# Ionically generated built-in equilibrium space charge zones – a paradigm change for lead halide perovskite interfaces

Gee Yeong Kim, Alessandro Senocrate, Davide Moia, and Joachim Maier[*]

Max Planck Institute for Solid State Research, Heisenbergstr. 1, 70569, Stuttgart, Germany

[*]E-mail: office-maier@fkf.mpg.de

**Methylammonium lead iodide (MAPI) is the archetype of the intensively researched class of perovskites for photovoltaics. Nonetheless, even equilibrium aspects are far from being fully understood. Here we discuss equilibrium space charge effects at the MAPI/TiO$_2$ and MAPI/Al$_2$O$_3$ interfaces, which are of paramount significance for solar cells. Different from the photovoltaic literature in which such built-in potentials are considered as being generated solely by electronic charge carriers, we will apply a generalized picture that considers the equilibrium distribution of both ionic and electronic carriers. We give experimental evidences that it is the ions that are responsible for the equilibrium space charge potential in MAPI, the reason being a pronounced ion adsorption at the contacts. The occurrence of equilibrium space charge effects generated by ionic redistribution has not been considered for photovoltaic materials and as such provides a novel path for modifying charge-selective interfaces in solar cells, as well as a better understanding of the behavior in mesoporous systems.**

In charge carrier containing systems any equilibrium interface carries an excess charge. Its magnitude depends on the nature of the phase (or grains) in contact, as well as on the control parameters defining the thermodynamic state. Only at a singular parameter set (point of zero charge or flat-band conditions) this charge disappears. The excess charge in the interface core and/or at the adjacent sides of the contact is compensated by space charges building up in the material under concern, which generally has finite ionic and electronic conductivities. As a consequence, all these charge carriers redistribute at the interface and hence lead to significant deviations of the ionic and electronic transport properties from the bulk value[1,2]. Charge carriers carrying an effective charge which is opposite to the charge to be compensated will be accumulated, while in the inverse case they will be depleted. These considerations necessarily apply to all mobile ionic and electronic charge carriers individually. The situation is characterized by the uniformity of the electrochemical potentials of both electrons and ions. This results in the uniformity of the electrochemical potential of the respective defects as well as of the chemical potential of the respective components. If only electronic variations occur, chemical equilibrium is not achieved. Moreover, space charge fields are very frequently generated by ionic interactions, in particular when ionic charge carriers are majority carriers (as it is the case for MAPI). Not only does this approach (termed nano-ionics) explain a great variety of "anomalies"[2-7], it has even led to paradigm changes in various research areas[1,4].

In view of the relevance of ion motion in halide perovskites[8,9], let us refer to two examples reported in the context of ionic conductors[5,6]: (i) At the LiF/TiO$_2$ interface a Li$^+$ transfer from LiF to TiO$_2$ leads to a positive excess charge (interstitial Li$^+$ and holes, $Li_i^\bullet$ and $h^\bullet$) on the TiO$_2$ side which is compensated by Li$^+$-vacancies ($V_{Li}'$) on the LiF side ("v-$i$-junction"). The accumulation of $V_{Li}'$ in LiF manifests itself in an enhanced ion conductivity, while the corresponding Li$^+$ excess on the TiO$_2$ side causes an accumulation of $Li_i^\bullet$ and is electronically perceived in terms of a transition from n- to p-type[6]. This is in sharp contrast to what would be expected for a Li component transfer (lithiation) or for an electroneutral bulk effect. (ii) As a second example of relevance, we consider composites of insulating oxides (Al$_2$O$_3$, SiO$_2$) and poor ion conductors[1-4,10,11]. Depending on the surface chemistry, cations or anions are preferentially adsorbed at the oxides surfaces leading to accumulation of the respective vacancies. Hence solid electrolyte composites such as LiI:Al$_2$O$_3$ or Ag-halides:Al$_2$O$_3$ show a strong enhancement of the cation conductivity as a consequence of Li$^+$ (Ag$^+$) adsorption and Li$^+$-vacancy (Ag) accumulation[10,12-15]. In AgCl:Al$_2$O$_3$, not only the increase of Ag$^+$ vacancy concentration, but also the impact of the so-established field on the electronic minority carriers was investigated and their interfacial conductivity was shown to be completely determined by the interfacial behavior of the ions ("fellow traveler effect")[15]. In fact all these effects are so well understood and powerful that purposeful admixing of surface-active second phases became an established strategy to optimize material properties. This method has been termed "heterogeneous doping" or "higher-dimensional doping", as opposed to the classic homogeneous or zero-dimensional doping relying on the presence of charged point-defects rather than charged interfaces[1,4]. (Note that the term heterogeneous doping is sometimes also used to refer to the different case of inhomogeneous classic doping *i.e.* to inhomogeneous distribution of zero-dimensional dopants) Not only has this experimental approach not been applied in the solar cell field, the phenomenon of built-in potentials being predominantly due to ionic effects as such has not been considered.

Here we will give evidence that analogous effects (related to local ionic defect formation energies) dominate the MAPI:$Al_2O_3$ and MAPI:$TiO_2$ interfaces. Given the importance of the MAPI:$TiO_2$ interface[16-21] in extracting the photo-generated electrons formed by illumination, this result is expected to be of significant influence for future interfacial engineering of perovskite solar cells. Using $Al_2O_3$ instead of $TiO_2$ is important in this study, particularly because $Al_2O_3$ is an insulator and redox-inactive[22], but also because the role of mesoporous $Al_2O_3$ played in early halide perovskite devices is not fully understood[22].

Our paper is structured as follows: (i) We will show that MAPI:$Al_2O_3$ composites (oxide nanoparticles dispersed in MAPI) exhibit a positive excess charge at the $Al_2O_3$ side of the contact and that this is the consequence of an ionic adsorption process. (ii) Then we will show that the situation in MAPI:$TiO_2$ is fully analogous. (iii) These findings will be corroborated by thickness-dependent conductance measurement in pure MAPI thin films on $Al_2O_3$ substrates. All the results are consistent with the generalized model used here to describe the space charge properties. (iv) At the end of the paper, we will address the nature of ionic adsorption and discuss implications for the solar cell function.

**Figure 1** sketches the generalized contact thermodynamics and its implications for the concentration distribution assuming a positive excess charge at the interface, as may be realized by cation adsorption. (We will indeed show later that such excess positive potential applies here). Accordingly, the ionic and electronic conductivities at the MAPI interface are expected to be drastically altered. For simplicity, we only display the expected variation for iodine vacancies ($V_I^\bullet$), conduction band electrons ($e'$) and holes ($h^\bullet$) in the diagram (**Fig. 1(b)**) as they are active in the conductivity experiments[8,9]. In the following, we demonstrate that exactly these predicted situations are met at the interfaces investigated here. It is well understood that iodine vacancies (similarly as oxide vacancies in oxide perovskites) form the majority defects in the bulk of MAPI. As such their concentration and hence the ion conductivity is rather insensitive with respect to the iodine partial pressure $P(I_2)$ (analogous to oxygen partial pressure dependence in ionically disordered oxides). For not too low $P(I_2)$, the electronic conductivity is of p-type and steeply increases with $P(I_2)$, while at very low $P(I_2)$ a switch to n-type conductivity is expected[8,9,23,24].

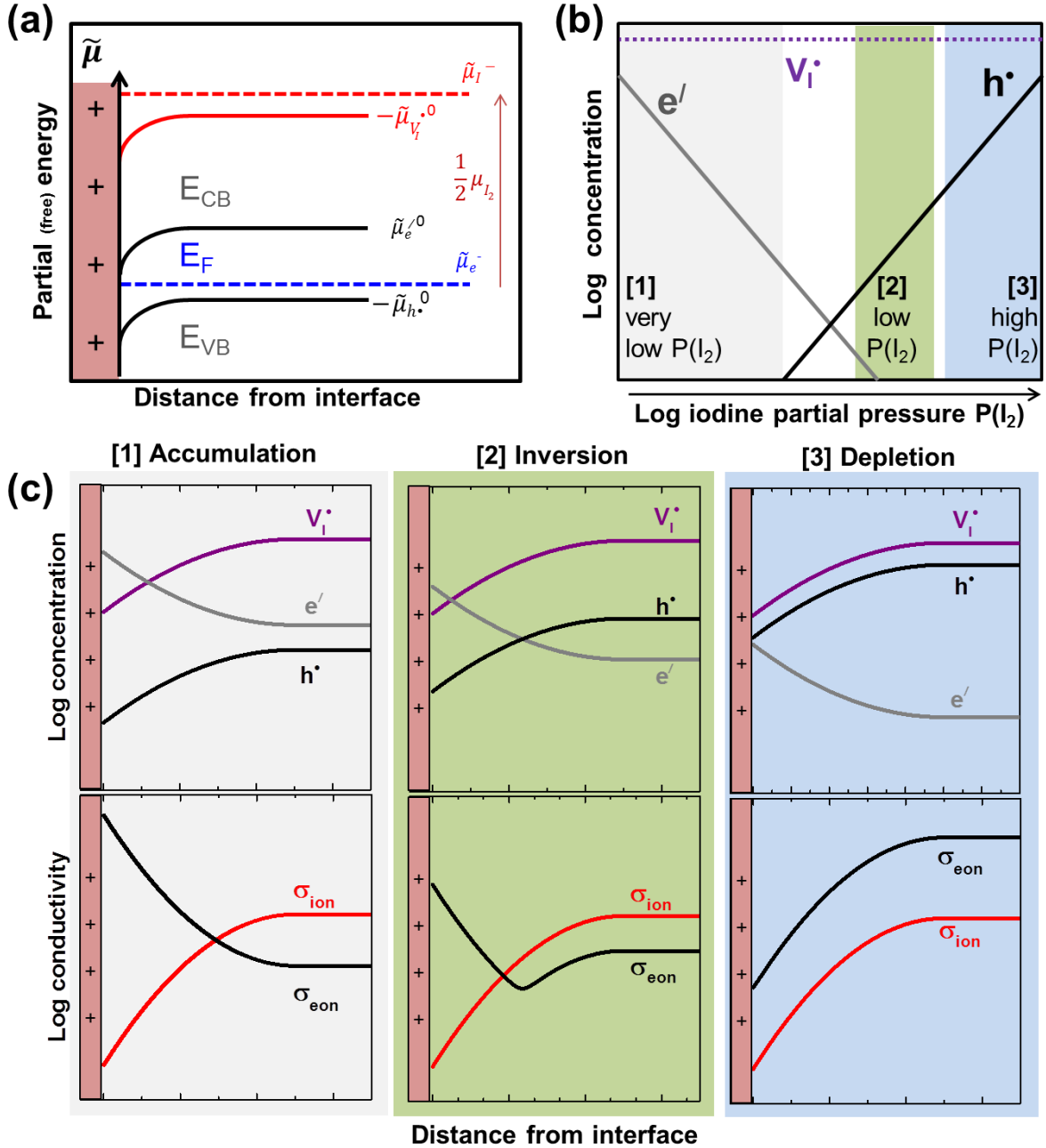

**Figure 1. Schematic diagram of charge carrier and conductivity distributions at a positively charged interface.** (a) Generalized energy level diagram with space charge equilibrium at the MAPI:Al$_2$O$_3$ (MAPI:TiO$_2$) interface. All the mobile electronic and ionic charge carriers are influenced by the common electric field ($\tilde{\mu}$: electrochemical potential, $\tilde{\mu}^0$: standard electrochemical potential). The coupling of ionic and electronic levels is determined by the iodine stoichiometry according to the well-known relation $\tilde{\mu}_{I^-} - \tilde{\mu}_{e^-} = \frac{1}{2}\mu_{I_2}$ [25]. (b) Charge carrier situation in MAPI at various $P(I_2)$ (i.e. $\mu_{I_2}$): [1] Very low $P(I_2)$ (not met in our measurements)

where conduction band electrons (n-type) are the dominating electronic charge carriers in the bulk and at the interface. [2] Low $P(I_2)$ (~Ar) where bulk is still p-type, but inversion to n-type occurs at the interface. [3] High $P(I_2)$ where holes are the dominant electronic carriers. [(c) Schematic concentration and conductivity ($\sigma_{eon}$, $\sigma_{ion}$) profiles for $V_I^\bullet$ (iodine vacancy), $h^\bullet$ (hole), and $e'$ (conduction band electron) at the interface between MAPI and the positively charged second phase at three different iodine partial pressures indicated as [1], [2] and [3] in (b). Mobilities of conduction band electrons and holes are assumed to be similar, but higher than for the ions. The compensation of the majority carrier ($V_I^\bullet$) could be intrinsic (*e.g.* $V_{MA}'$) or extrinsic (*e.g.* $Na_{Pb}'$ or $O_I'$).

If $P(I_2)$ is lowered, the ratio of hole and conduction band electron concentration is lowered and approaches the intrinsic situation, which is expected to be met at $P(I_2) < 10^{-9}$ bar.[8] The exact value of this cross-over partial pressure (see **Fig. 1(b)**) depends on the bulk vacancy concentration. Below this value the majority carrier is n-type. This situation (region [1] in **Fig. 1(b)**) is not met in our experiments (exposure to ultra-high vacuum would lead to decomposition). If the space charge potential is positive, both $\sigma_{eon}$ and $\sigma_{ion}$ are depressed due to depletion of holes and iodine vacancies in the adjacent space charge zones at high $P(I_2)$ (*i.e.* region [3] in **Fig. 1(c)**). On the other hand, at low $P(I_2)$ $\sigma_{eon}$ will eventually increase owing to the inversion from p to n-type even if the bulk is still p-type (*i.e.* region [2], **Fig. 1(c)**). $\sigma_{ion}$ is still depressed under these conditions.

All these features are observed in the composites of MAPI with $Al_2O_3$ and $TiO_2$ nanoparticles of 5-10 nm size (see **Fig. 2**). We measured and separated electronic conductivity ($\sigma_{eon}$) and ionic conductivity ($\sigma_{ion}$) by *d.c.* polarization measurements with interdigitated ion-blocking Au electrodes, a technique that has been already successfully applied to investigate charge transport properties in MAPI.[26] (The impact of Au electrode potentially showing interfacial effects is discussed in SI 1.2). As discussed in detail in SI 4.1, the fact that variations occur already at low volume fractions can be ascribed to the composite's microstructure and the small grain size of the particles, realizing conditions where coherent oxide networks (percolating paths) can be formed. In view of the defect chemical situation a Mott-Schottky behavior at the interface of interest is more probable than a Gouy-Chapman situation (see SI 1.1). The measured overall conductivity for such situation is given by

$$\sigma_m \simeq \sigma_\infty + \beta_L \varphi_A \Omega_A \lambda^* (Fu_n) c_{n,0} \left(2 \ln \frac{c_{n,0}}{c_{n,\infty}}\right)^{-1} \quad (eq.\ 1)$$

($\sigma_m$: measured overall conductivity, $\sigma_\infty$: bulk conductivity, $\beta_L$: percolation factor (~0.5), $\varphi_A$: volume fraction, $\Omega_A$: specific area (per unit volume), $F$: Faraday's constant, $u_n$: excess electron mobility, $c_{n,0}$ and $c_{n,\infty}$: excess electron concentration at interface and bulk, $\lambda^*$: width of Mott-Schottky space charge zone)[2]. The detailed calculation (given in SI 1.1) shows that all the results for the $Al_2O_3$ composites can be quantitatively explained by a space charge potential of about 740±60 mV (corresponding to an ion adsorption on the order of 1% of a monolayer, see SI 3.3). At this point it is noteworthy that we derive $\lambda^* \approx 20$ nm, a value that alleviates the percolation issue given the small particle size of 5-10 nm (see SI Fig. S11). While the derivation of the parameter values relies on literature data on mobilities[23,27] and densities of states[28] and on the assumption of a Mott-Schottky model, our conceptual conclusions are unaffected by such uncertainties.

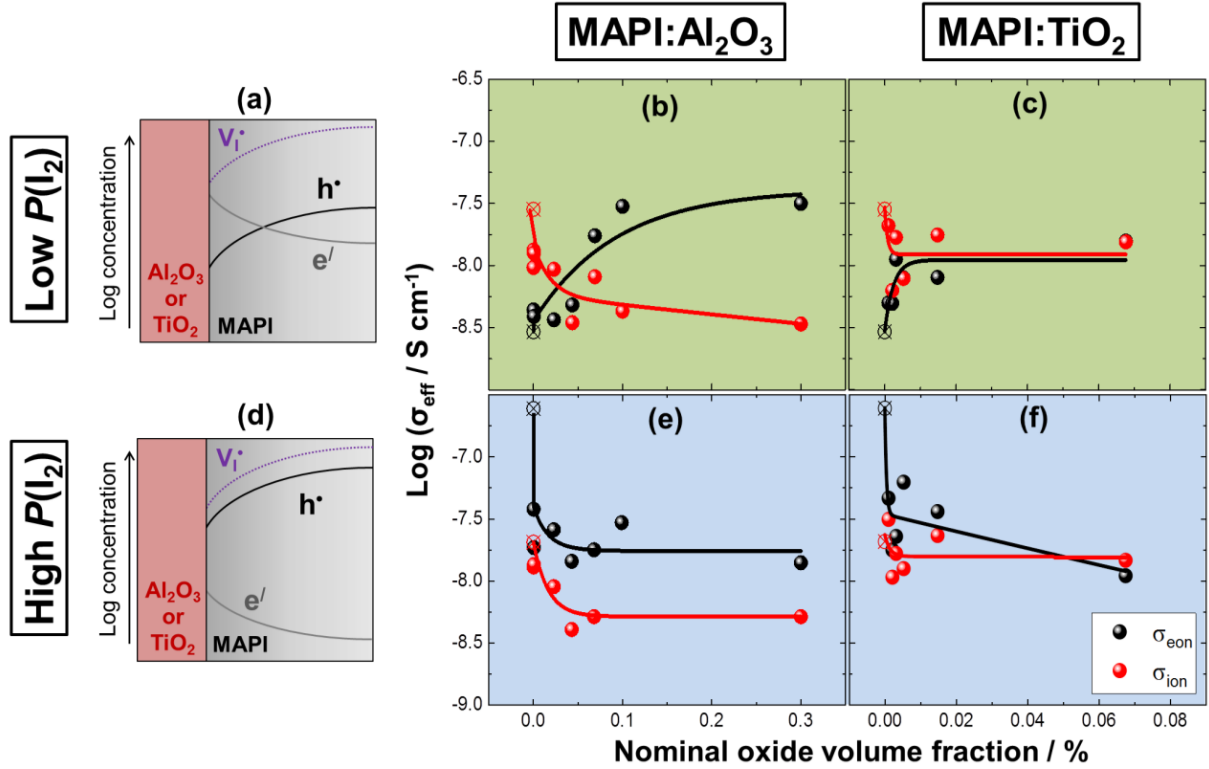

**Figure 2. Electronic ($\sigma_{eon}$) and ionic conductivities ($\sigma_{ion}$) of MAPI:$Al_2O_3$ and MAPI:$TiO_2$ composites.** (a) and (d) show a schematic of charge carrier ($V_I^\bullet$ (iodine vacancy), $h^\bullet$ (hole), and $e'$ (conduction band electron)) concentrations at the interface between $Al_2O_3$ (or $TiO_2$) and MAPI

for a positive charge on the oxide side. (b) and (c) refer to conductivities at low $P(I_2)$ and (e) and (f) refer to conductivities at high $P(I_2)$ ($\approx 10^{-6}$ bar) as a function of nominal volume fraction of oxide nanoparticles, extracted from *d.c.* galvanostatic polarization measurement at 40ºC. The volume fractions given in the figure are the ones in the precursor solution; according to ICP measurement, the real oxide volume fraction in the bulk is estimated to be higher by a factor of 10 (see SI 4.1). We plot effective conductivity (conductance corrected by macroscopic geometry) to take account of potential interfacial contributions by the Au-MAPI interface (see SI 1.2). The conductivity variations that a positive excess charge causes are very characteristic: $\sigma_{eon}$ is increased and $\sigma_{ion}$ decreased with increasing volume fraction at low $P(I_2)$. Both conductivities are reduced under high $P(I_2)$. Cross circle symbols indicate pristine MAPI sample (no oxide). Solid lines are guiding the eye.

To get further insight, we performed and evaluated conductivity measurements for various iodine partial pressures and hence various stoichiometries in MAPI (see SI 1.3). We now also include the results at high volume fractions. At these high values severe blocking effects occur that obscure the absolute $\sigma$-values, but as the microstructure stays constant on $P(I_2)$ variations, the results on the $P(I_2)$-dependence are reliable and even more representative of space charge effects. As far as the quantitative analysis is concerned, we refer to SI, with Table S2 giving the expected dependencies. The measured results are all consistent with the predictions as outlined in SI 1.3, and also suggest a transition to n-type at low $P(I_2)$. As the dependencies are the same as for the bulk, the agreement is a worthwhile consistency check, but does not provide us with more information.

All the above features can be also observed when $TiO_2$ is used as second phase (see **Fig. 2**): decrease of $\sigma_{ion}$ and $\sigma_{eon}$ if the ratio of p- to n-type conductivities is very large in the bulk, but characteristically an increase of the $\sigma_{eon}$ if this bulk ratio is not very large so that the space charge field will invert the conductivity type. The quantitative analysis gives a ~50 mV lower space charge potential for $TiO_2$ than for $Al_2O_3$, which is in fact not unexpected in view of the typically lower basicity (lower ionization degree of the oxygen ion) of $TiO_2$ (as reflected by the smaller $pH^*$ in water[29,30]). We also measured the conductivity of both $Al_2O_3$ and $TiO_2$ composites under light, where we do not see any accumulation effect under Ar. We briefly discuss this observation in the SI (Section 2).

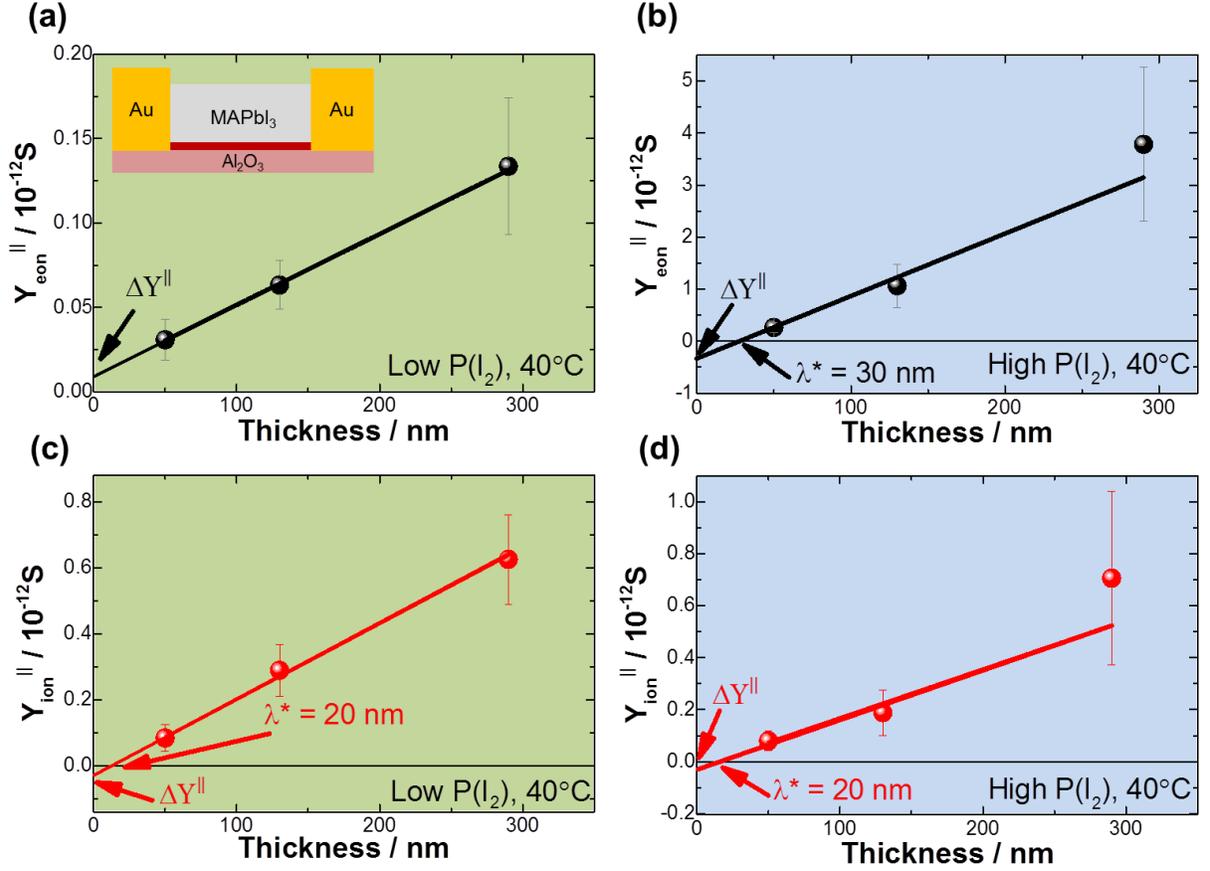

**Figure 3. Normalized electron and ion sheet conductance ($Y^{\parallel}$) plots as a function of MAPI thickness on $Al_2O_3$ substrates.** (a) Electronic and (c) ionic conductances at low $P(I_2)$ (Ar). (b) Electronic and (d) ionic conductances at high $P(I_2)$ ($\approx 10^{-6}$ bar). The *y*-intercept of the function $Y^{\parallel}(L)=\sigma_{\infty}L+\Delta Y^{\parallel}$, i.e. $\Delta Y^{\parallel}$, indicates excess conductance at the interface, and the slope corresponds to the bulk conductivity ($\sigma_{\infty}$). The fitting lines and error bars are obtained from a statistical analysis of 10 different samples. The inset sketches the measurement configuration; the red region indicates the space charge zone.

To corroborate the results obtained for the composites, parallel conductance measurements of MAPI thin films with various thicknesses deposited on $Al_2O_3$ substrates were performed to determine the charge accumulation and depletion effects[6,31-33]. If we plot the sheet conductance ($Y^{\parallel}$) as a function of thickness (L), we expect a straight line (for L >> screening length) with a slope corresponding to the bulk conductivity and an intercept corresponding to the interfacial contribution ($\Delta Y^{\parallel}$). As shown in **Figs. 3(b)** and **(d)**, we find that for high $P(I_2)$ as well as for low $P(I_2)$ (Ar) the bulk ionic and electronic sheet conductances increase with film thickness more or less linearly. In all cases, values of the bulk conductivities derived from the slopes are close to the ones measured for bulk MAPI (see SI 1.1 for details). In the case of the ionic conductance the intercept is below zero indicating a depletion effect under both conditions. The electronic conductance shows a negative intercept (*i.e.* depletion) at high $P(I_2)$, but characteristically a positive intercept indicating accumulation under low $P(I_2)$ [see **Fig. 3(a)**]. All these results are in agreement with the composite values. Only the absolute value of the space charge potential is lower by 120±60 mV than for the composites, *i.e.* about 620±80 mV (see detailed calculation SI 1.1). Yet again, this lower activity is expected from the surface chemistry of $Al_2O_3$ ($\gamma$-$Al_2O_3$ nanoparticles in the composites versus $\alpha$-$Al_2O_3$ substrates) and previous experience in heterogeneous doping[2]. It is also meaningful in the case of depletion to pay attention to the intersection with the thickness-axis. In the simplest approximation this should yield the width of the depletion zone. The so-obtained value of the Mott-Schottky screening length ($\lambda^*$) is 20±10 nm. Based on this results, we obtain a Debye length of ~2±1 nm, corresponding to a bulk defect concentration on the order of 1000 ppm (SI 1.1). From this, we can derive a vacancy mobility of ~$10^{-8}$ $cm^2$/Vs (see SI 1.1 for details).

Unfortunately, $TiO_2$ could not be used as a substrate to perform analogous experiments. However, we have seen that (i) the space charge results of the $TiO_2$ composites coincide with the ones for the $Al_2O_3$ composites, and (ii) the adsorptive behavior of $TiO_2$ concerning the relevant ions ($Pb^{2+}$, $I^-$, $MA^+$) in solution coincides with the behavior of $Al_2O_3$ as discussed below (see details in SI Section 3). This similarity together with the fact that $Al_2O_3$ is redox-inactive (*i.e.* the $Al_2O_3$/MAPI interface is determined by an ionic redistribution effect) leads to the conclusion that an analogous (*i.e.* ionically dominated) interfacial situation applies to MAPI/$TiO_2$. Owing to the similarity to the composite electrolytes (LiI:$Al_2O_3$, AgCl:$Al_2O_3$, CuCl:$Al_2O_3$, or TlCl:$Al_2O_3$), where the cations ($Li^+$, $Ag^+$, $Cu^+$, $Tl^+$) are adsorbed at the oxide's surface[2,10], we may expect our effects also to be caused by cation adsorption.

While it is clear that, in MAPI, anions are mobile and cations only show very small conductivities[9], this does not preclude such an adsorption, as boundary redistributions within few nm do not require significant conductivities. More importantly, such cation enriched adsorption layers can simply form during the liquid-state synthesis. We note also that segregation of positively charged iodine vacancies could be a plausible mechanism, as this process would be equivalent to a preferential excess of both cations and could occur kinetically more easily. In the literature one finds experimental reports on Pb-$TiO_2$ interaction[34], as well as calculations claiming that $PbI_2$-terminated MAPI is more stable than the MAI-terminated MAPI in contact with anatase[35]. While these results are not necessarily meaningful for our discussion, for the case of $\gamma$-$Al_2O_3$ there are explicit reports about a significant adsorption of Pb(II)[36,37]. Even more conclusive in this context are adsorption studies that we performed for $Al_2O_3$ and $TiO_2$ surfaces with respect to $Pb^{2+}$, $MA^+$ and $I^-$ ions in DMSO solution. They are described in greater detail in SI Section 3. Results from ICP, Zeta-potential, STEM-EDX, and NMR measurements indicate $Pb^{2+}$ adsorption rather than $MA^+$ or $I^-$ adsorption. When the oxide particle is removed from a DMSO solution containing $MAI+PbI_2$, it exhibits a surface layer of $Pb^{2+}$ with $I^-$ as counter ion, corresponding to a coverage of about 30% of a monolayer (detail estimate is given in SI 3.3). This adsorption behavior can certainly not be accurately translated into adsorption properties of the solid situation, but should capture the chemical trend. To stress again the point that is important for our interpretation: $TiO_2$ behaves exactly the same as $Al_2O_3$ in these studies.

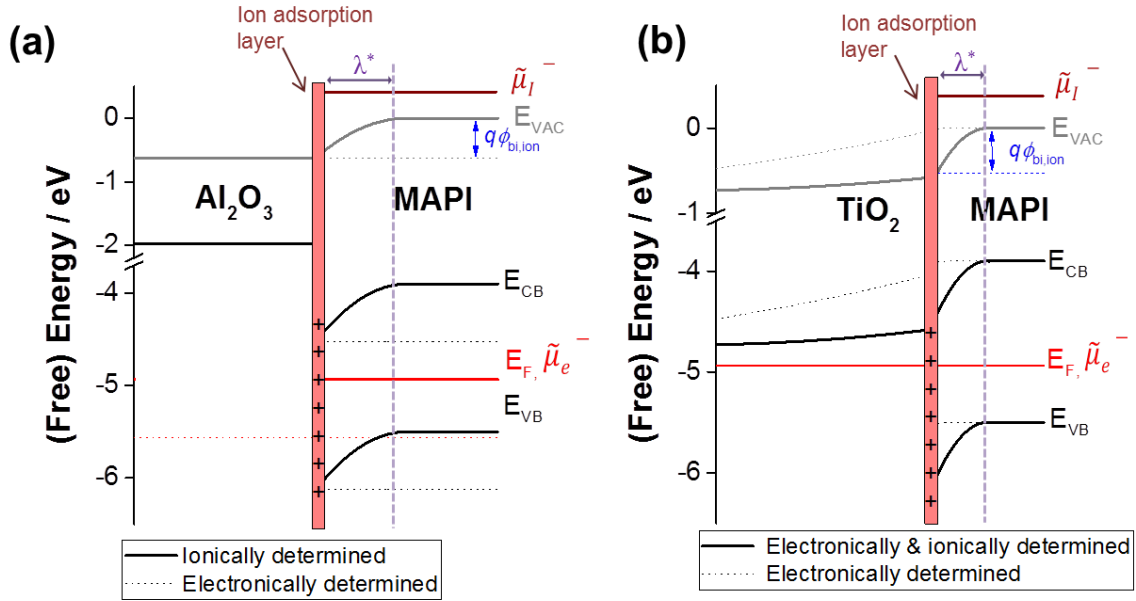

**Figure 4. (Free) Energy band diagram at the interface between MAPI and oxide.** (a) $Al_2O_3$/MAPI and (b) $TiO_2$/MAPI interfaces. The hypothetical case of an electronically determined space charge situation is referred to by dashed lines and the actual ionically determined space charge situation is referred to by solid lines. The electronic built-in potential is defined as $\phi_{bi,eon} = \phi_{MAPI} + \phi_{TiO_2}$ and the ionic built-in potential ($\phi_{bi,ion}$) can be obtained from our calculated values. (For the work function (4.2 eV for $TiO_2$ and 4.9 eV for MAPI) and band gap values we refer to [28,38-40])

So far we have concluded that the built-in space charge on the MAPI side is ionically dominated [see **Fig. 4(a)**] when in contact with $Al_2O_3$ and $TiO_2$. Unlike $Al_2O_3$, on the $TiO_2$ side electronic effects do also contribute to the overall built-in potential [**Fig. 4(b)**]. The significant impact of ion distribution conforms to the experience that we have with the Solid State Ionics of composites, and is also expected as ionic defects are the bulk majority carriers in MAPI. For the same reason the generally observed insensitivity of conductivity effects with respect to MAPI stoichiometry or sample variation (*e.g.* microstructure) is not surprising. Nevertheless, this point has to be investigated more systematically in the future.

As a next step, we discuss the impact of ionically generated space charges on solar cell devices. Needless to say, space charges influence the electron/hole extraction and affect the recombination dynamics at such interfaces[41-43], thus being of paramount importance for photovoltaics. Surface recombination is considered to be the dominant recombination loss in high efficiency halide perovskite solar cells[44,45], and the charge distribution at interfaces a key parameter[41]. One example is a recent work[46] which stresses the importance of interfacial capacitances in influencing recombination and injection. These capacitances can be modified by varying the space charge layer width, which according to our finding is determined by the ion adsorption. In silicon solar cells, "high-$\kappa$" dielectric such as $Al_2O_3$, $HfO_2$ and $ZrO_2$ films have been used to improve solar cell efficiency by providing effective surface passivation. This effect was attributed to a hydrogen-induced passivation mechanism as a consequence of a negative surface charge[47,48].

While the sign of the space charge effect on the MAPI side conforms to previous literature reporting electron accumulation and accounting for electronic effects only[21], the central role of the ionic properties in the determination of the space charge potential is a novel conception. Specifically, it implies that the fraction of built-in potential of these interfaces dropping across the perovskite layer [$q\phi_{\text{bi,ion}}$ in **Fig. 4(b)**] is largely determined by the ionic interaction with the oxide surface. In the example shown in **Fig. 4(b)**, the ionically dominated space charge potential in MAPI is larger than what would be expected considering electronic effects only. For cases where Shockley-Read-Hall recombination via deep traps at this interface is dominating, the increased band bending at the MAPI side should be beneficial to device performance (due to a decrease of the local minority carrier concentration). Note that typical values of ionic carrier concentrations are higher than the concentration of photo-generated carriers even under light[23,49]. This is even more so, as the ionic carrier concentration may be significantly increased by illumination, which, in turn, complicates the space charge picture (see SI Section 2).

Another point worth noting is that, according to our experiments, MAPI undergoes a p-to-n transition at the $Al_2O_3$ (or $TiO_2$) interface due to ion adsorption. For devices where the hybrid perovskite is infiltrated in mesoporous oxide films with pore size in the order of 20-50 nm, our results suggest that considerable depletion of iodine vacancies and of holes (p- to n- type transition) may occur throughout the perovskite phase. This would have important consequences in terms of potential distribution between the perovskite that is infiltrated in the oxide and the one forming the capping layer. It would also influence the charge transport and recombination dynamics in the mesoporous film.

A further implication is that such effects can -given the moderate adsorption value of ~1% monolayer- be altered in either direction by surface modification, by doping $TiO_2$ or by strategically looking for alternative phases replacing $TiO_2$. In this way, band alignment or band bending can be varied to control the collection and recombination of photo-generated charges, strikingly without the need to vary the bulk properties of the contact phases. Stoichiometry control (as well as doping) on MAPI may also alter the ion adsorption behavior[50,51]. While recent interfacial engineering studies on halide perovskites showed that recombination could be suppressed by inserting self-assembled monolayers between the electron transfer layer and perovskite layer, changes in the ionic surface properties have only been scarcely invoked as a possible explanation for the observed trends[39-41].

In conclusion, our findings give strong evidences for ionically determined equilibrium space charge potentials forming at the MAPI/$Al_2O_3$ and also MAPI/$TiO_2$ contacts. These are attributed to a positive ionic excess charge at the interface, most probably due to $Pb^{2+}$ adsorption. The so-formed positive space charge, which should also apply to other lead halide perovskites, leads to a local depletion of holes and iodine vacancies, and an accumulation of conduction band electrons. Owing to the high ionic charge carrier concentration involved (exceeding typical concentrations of photo-generated electronic carriers), these distribution phenomena are expected to be of relevance even under illumination/in devices under operation. The significance of this result is enormous, given the fact that in photovoltaics such equilibrium space charge phenomena are generally discussed solely in terms of electronic effects. The role of the ionic properties on the space charge potentials as discussed here enables completely novel approaches of interface engineering with respect to potentially improve solar cell efficiency.

52  Im, J.-H., Lee, C.-R., Lee, J.-W., Park, S.-W. & Park, N.-G. 6.5% efficient perovskite quantum-dot-sensitized solar cell. *Nanoscale* **3,** 4088-4093 (2011).
53  Jeon, N. J. *et al.* Solvent Engineering for High-Performance Inorganic-Organic Hybrid Perovskite Solar Cells. *Nat. Mater.* **13,** 897 (2014).
54  Gillespie, L. J. & Fraser, L. H. D. The Normal Vapor Pressure of Crystalline Iodine. *Journal of the American Chemical Society* **58,** 2260-2263 (1936).

**Acknowledgements**

The authors are very grateful to Kersten Hahn, Julia Deuschle, and Chen Li (Stuttgart Center for Electron Microscopy, led by Peter van Aken) for performing the STEM, EDX measurements and FIB. The authors are also thankful to Kathrin Müller (Interface Analysis group in the Max Planck Institute for Solid State Research, led by Ulrich Starke) for XPS measurements. The authors are grateful to Helga Hoier for XRD measurements, Igor Moudrakovski for NMR measurement, Achim Güth (Nanostructuring Lab in the Max Planck Institute for Solid State Research, led by Juergen Weis) for his help with electrode deposition and Samir Hammoud for ICP measurement. The authors also wish to thank Jelena Popovic and Feixiang Wu for helpful discussion.
**Experimental Methods**

*Preparation of MAPbI$_3$ thin-films for electrical measurements*

MAI was obtained using the reported procedure by Im *et al.*[52] An equimolar solution of MAI and PbI$_2$ (1.5 M) in DMSO was prepared and spin-coated on polished sapphire (0001) substrates, previously equipped with interdigitated Au electrodes (2 μm spacing, 100 μm length). Prior to the spin-coating step, substrates were cleaned with an O$_2$-plasma treatment, in order to remove organic impurities and enhance the hydrophilicity of the surface. For the composites, commercial nanoparticle of Al$_2$O$_3$ (Sigma Aldrich, 544833), or TiO$_2$ (Sigma Aldrich, 637254) are dispersed in the equimolar MAI + PbI$_2$ solution at different volume fractions with respect to the solid MAPI. During spin-coating, a drop of chlorobenzene was used to induce rapid crystallization,[53] and the sample was later annealed for 5 minutes at 100 °C. For the thickness dependent experiments, we deposited MAPI layer on Al$_2$O$_3$ substrate with different thickness by changing the concentration of MAPI. The thickness of film was confirmed by profilometer (Bruker Dektak) and AFM (Bruker Dimension Icon) measurements. All procedures were carried out in an Ar-filled glovebox (O$_2$ < 0.1 and H$_2$O < 0.1 ppm)

*D.c.-galvanostatic polarization*

*D.c.* polarization experiments were performed by using a current source (Keithley model 220) and by monitoring the voltage change with a high impedance electrometer (Keithley model 6514). Measurements were carried out in the dark and under light illumination and by accurately controlling temperature and atmosphere over the sample (oxygen content and humidity were monitored using appropriate sensors). As the light source, a xenon arc lamp was used, and its intensity was calibrated using a power meter (Oriel). We varied the film thickness by controlling the concentration of the precursors in the solution and measured the conductance via Au ion-blocking electrodes in an interdigitated arrangement. Au electrodes were considered as ideally selective electrodes. The possible presence of interfacial effects, however, is not expected to substantially change the picture (see SI 1.2).

*Iodine partial pressure dependence*

To control $P(I_2)$ over the samples, argon was used in a container with solid iodine chips, kept in a thermostat at a fixed temperature (always below room temperature, between −40°C and 4°C). The iodine partial pressure was assumed to correspond to the equilibrium pressure of iodine at the thermostat temperature, which was calculated based on a published equation[54]. Similar values are obtained by estimating $P(I_2)$ purely from thermodynamic considerations, starting from Gibbs free energy of sublimation of solid iodine.

*ICP measurement*

The ICP-OES measurements were performed by using SPECTRO CIROS. The intensity of measured each line is then compared to previously measured intensities of known concentrations of the elements, and their concentrations are then computed by interpolation along the calibration lines. We dissolved MAI and $PbI_2$ in DMSO and then immersed oxide particles into a solution. We quantify the Pb amount by ICP in a control solution without nanoparticles and compare it with the concentration left after adding $Al_2O_3$ or $TiO_2$ nanoparticles separating them by centrifuge.

*Zeta Potential measurement*

An electroacoustic DT-1200 spectrometer (Dispersion Technology, Inc., Quantachrome) was used to measure the colloidal vibrational current (CVI) at 3 MHz from which the Zeta potential was calculated. We dissolved salts (MAI, $PbI_2$, $Pb(NO_3)_2$, KI) in DMSO adding the oxide nanoparticles ($TiO_2$ and $Al_2O_3$) at different weight fractions. The measurements were performed at room temperature.

*STEM & EDX measurement*

Samples for the investigations were gently dispersed onto holey carbon Cu grids, by dropping the precursor solution containing nanoparticles on the grids in the glove box. After evaporation of the solvent, HR-TEM was performed at 200 and 80 kV with an advanced TEM (JEOL ARM200F, JEOL Co. Ltd.), equipped with a cold field-emission gun and a CETCOR image corrector (CEOS Co. Ltd.).

# Supplementary Materials for

# Ionically generated built-in equilibrium space charge zones – a paradigm change for lead halide perovskite interfaces

G. Y. Kim, A. Senocrate, D. Moia, and J. Maier

# 1. Evaluation of space charge parameters

## 1.1. Ideally selective electrodes

In this section we consider Au to be an ideally selective electrode (no interfacial effect). It is fully blocking for ions, but entirely reversible for electrons. Furthermore, we assume no interfacial resistance. In 1-2, we discuss with this last assumption.

### 1.1.1. Space charge contributions (Mott-Schottky vs. Gouy-Chapman type situation)

In principle the space charge zone between MAPI and oxides can refer to a Mott-Schottky or a Gouy-Chapman-type situation. For the evaluations, we have to distinguish between the transport along (parallel) or across (vertical) the space charge zones. When the majority charge carrier is mobile and Poisson-Boltzmann equation applies for all relevant carriers, Gouy-Chapman (GC) profiles are obtained for the space charge zones (see detail Fig. S1). On the other hand, the Mott-Schottky (MS) model is relevant in the case where one of the majority carriers (typically the impurity) is immobile and its concentration is spatially constant and the compensating majority carrier (we assume $V_I^\bullet$ with $[V_I^\bullet] \equiv c_v$) is depleted. In the MS case the screening is smaller and the width of the space charge zone wider. In the GC-case the Debye length $\lambda = \sqrt{\dfrac{\varepsilon\varepsilon_0 RT}{2F^2 c_{v,\infty}}}$ (or better its double value $2\lambda$) is the relevant screening length ($\infty$ indicates bulk), while in the MS-case the relevant screening length is the Mott-Schottky length $\lambda^* = \lambda\sqrt{\dfrac{4F}{RT}\left|\Delta\phi_0^{MS}\right|}$ which exceeds $\lambda$ depending on the value of the space charge potential $\left(\Delta\phi_0^{MS}\right)$.

Now let us consider the case of n-accumulation (accumulation of conduction band electrons) in a MS situation. The effective carrier conductivities in the space charge zones that have to be multiplied with $\lambda^*$, to give the overall conductance contribution, are given in Refs[1-3]. The ratio shows that in the MS-case the conductivity effect for a measurement along the interface $\left(\Delta\sigma_m^{\|}\right)$ is distinctly higher than in the GC-case for a given space charge potential (given $c_n$-enhancement). We use for the MS-case

$$\Delta\sigma_{m,n}^{\|MS} \simeq \Omega\lambda^*(Fu_n)c_{n,0}(2\ln\frac{c_{n,0}}{c_{n,\infty}})^{-1} \qquad (eq.\ 1)$$

and for the GC-case

$$\Delta\sigma_{m,n}^{\|GC} \simeq \Omega u_n 2\lambda\sqrt{c_{n,0}c_{n,\infty}} \qquad (eq.\ 2)$$

where for the composite $\Omega$ can be replaced by $\beta_L\varphi_A\Omega_A$. From the eqs 1 and 2, it follows that

$$\left|\frac{\Delta\sigma_{m,n}^{\|MS}}{\Delta\sigma_{m,n}^{\|GC}}\right| = \frac{1}{2}\frac{1}{\sqrt{\ln\frac{c_{n,0}}{c_{n,\infty}}}}\sqrt{\frac{c_{n,0}}{c_{n,\infty}}} \qquad (eq.\ 3)$$

($\Delta\sigma_m^{\|}$: overall conductivity contribution from the space charge zone, $\varepsilon$: dielectric constant, $\varepsilon_0$: permittivity of vacuum, $\sigma_\infty$: bulk conductivity, $\Omega$: interfacial area per volume, $\beta_L$: percolation factor (~0.5), $\varphi_A$: volume fraction, and $\Omega_A$: specific area, i.e. area per volume of the oxide particles (A), F: Faraday's constant, R: gas constant, $u_n$: excess electron mobility, $c_{n,0}$: excess electron carrier concentration at interface, $c_{n,\infty}$: excess electron carrier concentration in the bulk, $c_{v,\infty}$: vacancy concentration in the bulk, $\lambda^*$: width of the space charge zone in the MS-case). Eq. 3 means that MS requires less drastic boundary values than GC to explain a given $\Delta\sigma_m^{\|}$ (overall conductivity contribution from the space charge zone). Apart from this point, it is the depletion of the mobile iodine vacancies as majority carriers and the low mobility of the counter carrier (impurities or frozen native ionic defects, *e.g.* $V_{MA}^{/}$, $I_i^{/}$, etc.) that let us favor a MS-model. It is important to mention that this is a tentative procedure, and the reality may lie in between GC and MS situation.

### 1.1.2. Evaluation of thickness-dependent experiments

We performed parallel conductance measurements of MAPI thin-films with various thicknesses deposited on $Al_2O_3$ substrates to determine the excess charge accumulation and depletion effects. The total sheet conductance $Y^{\|}$, in this case, can be expressed as a function of thickness L by (see details in Ref [4-8]),

$$Y^{\|}(L) = \sigma_{\infty} L + \Delta Y^{\|} \qquad \text{(eq. 4)}$$

If the substrate conductivity can be neglected, $\Delta Y^{\|}$ corresponds to the interfacial space charge contribution in the MAPI film. For $L \gg \lambda^*$, a linear dependence of $Y^{\|}$ with respect to thickness is observed (*i.e.* $Y^{\|} = \Delta Y + \sigma_{\infty} L$) with $\Delta Y^{\|}$ being the intercept and the bulk conductivity ($\sigma_{\infty}$) being the slope. The intercept is positive in the case of an accumulation (electron: n) and negative in the case of depletion (hole: p, vacancy: v). We indeed see an accumulation effect for $\Delta Y_{eon}^{\|}$ under Ar-atmosphere, *i.e.* a positive $\Delta Y_{eon}^{\|}$, that can be ascribed to the inversion from p-type to n-type; on the other hand, as expected, we observe a depletion effect (negative intercept) for $\Delta Y_{ion}^{\|}$. The depression of $\Delta Y_{eon}^{\|}$ and $\Delta Y_{ion}^{\|}$ at higher $P(I_2)$ is in line with the p-type conductivity and iodine vacancy conductivity undergoing depletion of their respective charge carriers (see Fig. 3(b) in the main text). This thickness dependence of pure MAPI on $Al_2O_3$ substrates measurement results agree with the composite results.

### 1.1.2.a. Accumulation of excess electrons

Here we refer to the inversion regime (low $P(I_2)$, Ar) where the excess conductance due to accumulation of electrons at the interface is given by:

$$\Delta Y_n^{\|} = \lambda^* \frac{1}{2\ln\frac{c_{n,0}}{c_{n,\infty}}} \sigma_{n,\infty} = \lambda \frac{1}{\sqrt{\ln\frac{c_{n,0}}{c_{n,\infty}}}} (\frac{c_{n,0}}{c_{n,\infty}}) \sigma_{n,\infty} \qquad \text{(eq. 5)}$$

### 1.1.2.b. Depletion of holes and vacancies

The simplest way to address depletion effects is to neglect the conductance contribution within the space charge zone ($\lambda^*$). Then

$$\Delta Y_p^{\|} = (L - \lambda^*)\sigma_{p,\infty} \qquad \text{(eq. 6)}$$

and

$$\Delta Y_v^{\|} = (L - \lambda^*)\sigma_{v,\infty} \qquad \text{(eq. 7)}$$

This equation also shows that we then expect an intersection of the lines with the L-axis at L=$\lambda^*$. A mean value of 20±10 nm is obtained when considering all thin-film results in this study for the $Al_2O_3$ substrate.

### 1.1.3. Evaluation of the composites

**1.1.3.a. Absolute value of space charge conductivity: accumulation of excess electrons**

Again we find a striking conductivity enhancement if the electronic conductivity for MAPI:$Al_2O_3$ or MAPI:$TiO_2$ composites is measured under Ar (where we expect inversion from p-type to n-type to occur and n-type conduction dominating at the interface). Following previous work on composite electrolytes[1] the overall measured conductivity (MS situation, impurity level $[V_I^\bullet]_\infty = [A^/]_\infty$) can be written as

$$\begin{aligned}
\sigma_m &= (1-\varphi_A)\beta_\infty \sigma_\infty + \beta_L \varphi_A \Omega_A \lambda^* (Fu_n) \frac{c_{n0}}{2\ln\frac{c_{n0}}{c_{n\infty}}} \\
&\simeq \sigma_\infty + \beta_L \varphi_A \Omega_A \sqrt{\frac{\varepsilon\varepsilon_0 RT}{2c_{v\infty}}} u_n \frac{c_{n0}}{\sqrt{\ln\frac{c_{n0}}{c_{n\infty}}}} \qquad \text{(eq. 8)} \\
&\simeq \sigma_\infty + \beta_L \varphi_A \Omega_A \lambda \frac{c_{n0}}{c_{n\infty}} \ln\frac{c_{n0}}{c_{n\infty}} \cdot \frac{\sigma_{n0}}{\sqrt{\ln\frac{c_{n0}}{c_{n\infty}}}} \\
&\simeq \sigma_\infty + \beta_L \varphi_A \Omega_A \lambda \exp(\frac{F\Delta\phi}{RT}) \frac{\sigma_{n0}}{\sqrt{F\Delta\phi_0/RT}}
\end{aligned}$$

We assume that the particles give rise to percolating space charge conduction between the two electrodes and fully contribute (parallel switching, $\beta_L \approx 0.5$). The volume fraction of the interfacial layers is expressed by the oxide volume fraction ($\varphi_A$), their surface-to-volume ratio ($\Omega_A \simeq r_A/3$) and the MS length ($\lambda^*$) as layer thickness. (cf. Fig. S1).

## 1.1.3.b. Absolute values of space charge conductivity: depletion of holes and vacancies

Consistently with our model, we find that $\sigma_{m,eon}$ (measured electronic conductivity) is depressed at high iodine partial pressure where we find the (depleted) holes. A depression is also found for $\sigma_{m,ion}$ (measured ionic conductivity) under all conditions due to depleted vacancies. For a rough evaluation we refer to Fig. S1. We assume that the bulk is partially blocked by the dense oxide layers. A quick estimate uses a series switching between bulk and blocking layer, *i.e.* $\rho_m = \rho_\infty + \beta_l \varphi_l \rho_l$. The effective conductivity of the overall layer is then determined by the passage through the blocking particle arrangement. The volume of the passage (dense oxide layer) is 1/8 ~ 1/2 of the blocking layer volume ($2r_A \times$ area covered, see details below Fig. S1). The factor 1/2 applies if the total free volume between the densely packed particles is active; the factor of 1/8 applies if a cylinder of the width of the bottleneck is decisive. Taking a factor of 1/5 (between 1/2 ~ 1/8), one estimates (*cf.* Fig. S1) the measured effective resistivity as

$$\rho_m \simeq \rho_\infty + 3\varphi_A \frac{1}{\sigma_{sc}^{\parallel}} \qquad (eq.\ 9)$$

The first term describes the resistivity attributable to the bulk ($\rho_\infty$), and the last term represents the contribution from the space charge zone. The parameter $\beta_l$ refers to the proportion of pathways contributing to $\sigma_m$; $\varphi_l$ is the volume fraction and $\sigma_{SC}^{\parallel}$ is the conductivity in the conductive passage.

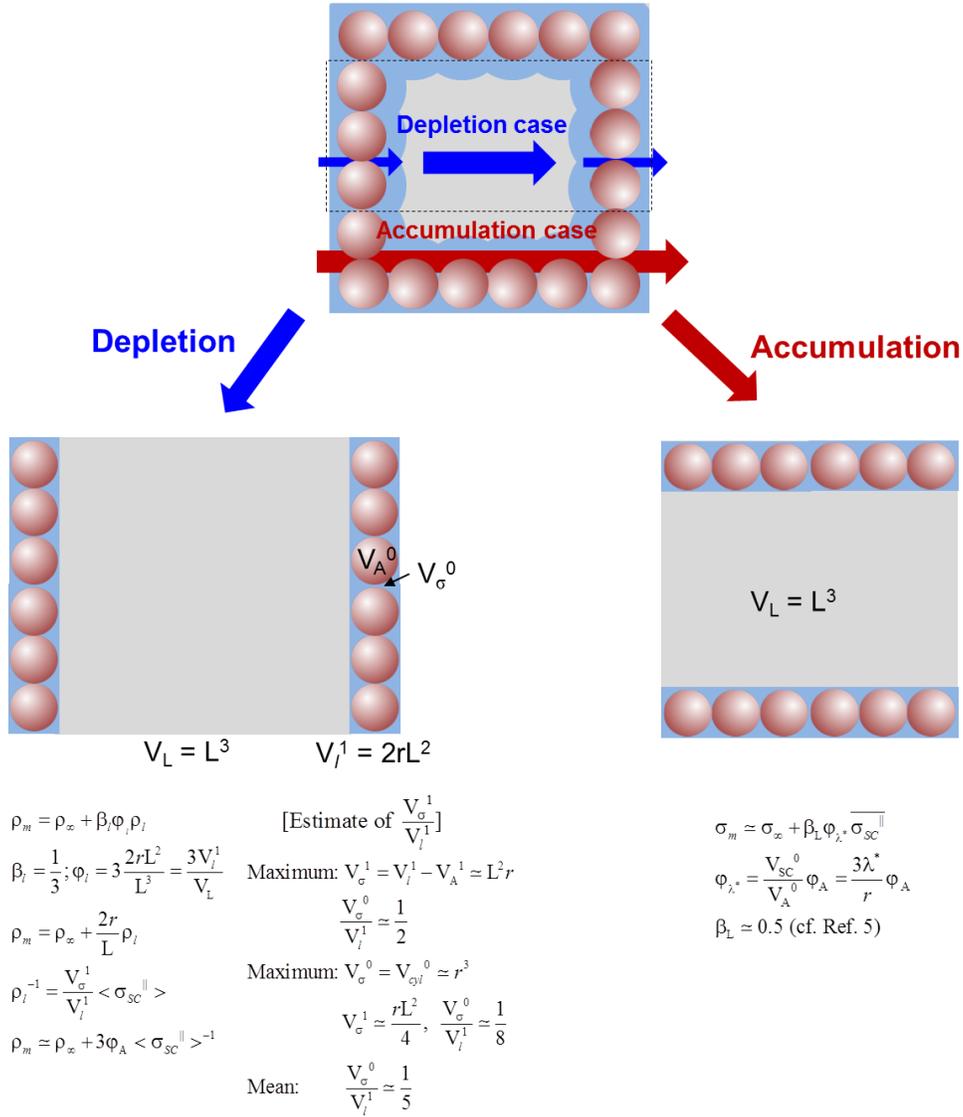

**Figure S1.** A schematic view of the composite for estimating accumulation and depletion effects. The length of the grain of MAPI is $L$ and the diameter of the particle is $l$.

As the width of the passage is smaller than the particle radius, the effective conductivity in the passage $\sigma_{sc}^{\parallel}$ is a space charge conductivity and not far from $\sigma_{p0}$ or $\sigma_{v0}$ if $r \ll \lambda^*$. Of course this gives only an upper limit for $\rho_m$, as any deviation from this morphology leads to significant conductive leakages. Using the above eq. 9, the $\sigma_{SC}^{//}$ value is found to be around $10^{-10} \sim 10^{-12}$ Scm$^{-1}$.
7

### 1.1.4. Parameter evaluation

If $\sigma_\infty$ is evaluated from the slopes, the results (given in Table S1) agree well with the bulk conductivities derived from earlier *d.c.* conductivity measurements on pure MAPI (Ar: $\sigma_{eon,\infty} = 3\times10^{-9}$, $\sigma_{ion,\infty} = 3\times10^{-8}$, high $P(I_2)$: $\sigma_{eon,\infty =} 4\times10^{-7}$, $\sigma_{ion,\infty} = 2\times10^{-8}$ Scm$^{-1}$)

**Table S1**. Calculated values for bulk conductivity extracted from the slope of conductance measurement on various thickness.

|  | $\sigma_{eon,\infty}$ / Scm$^{-1}$ | $\sigma_{ion,\infty}$ / Scm$^{-1}$ |
|---|---|---|
| **Ar** | 9×10$^{-9}$ | 4×10$^{-8}$ |
| $P(I_2) = 10^{-6}$ **bar** | 1×10$^{-7}$ | 1×10$^{-8}$ |

As already mentioned, the evaluation of the intersection of $Y^\|(L)$ in the case of thin films with the L-axis yields

$$\lambda^* \simeq 20\pm10 \text{ nm} \simeq \lambda\sqrt{\frac{4F\Delta\phi_0}{RT}}$$

From the ratio of the absolute values for $\Delta Y^\|$ (eq. 1) and $\Delta\sigma_m^\|$ (eq. 5) in the case of the enhancement effects, we derive for the differences in the space charge potentials a value of:

$$\Delta\phi_0^C - \Delta\phi_0^F \simeq 120\pm60 \text{mV}$$

($\Delta\phi_0^F$ : space charge potential from film data, $\Delta\phi_0^C$ : space charge potential from composite)
Note that combining Eq 1 and 5 leads to $\lambda$ canceling out.

Similarly from the composite values for n-accumulation in the Al$_2$O$_3$ and TiO$_2$ composites we find that $\Delta\phi_0$ is about 50 mV lower for TiO$_2$ than for Al$_2$O$_3$.

From the depression results, *i.e.* from Eq. 9, we can derive $\sigma_{p,0}$ and $\sigma_{v,0}$ and then since we know $\sigma_{p,\infty}$ and $\sigma_{v,\infty}$, the space charge potentials for the composites $\Delta\phi_0^C$ (and from this $\Delta\phi_0^F$ according to $\Delta\phi_0^F = \Delta\phi_0^C$ -120 mV). Yet as mentioned, in this case we can only get a lower limit for $\Delta\phi_0$ ($\Delta\phi_0 > 300$ mV). The results obtained from the absolute values of $\sigma_{m,eon}$ (Ar) and $Y_{eon}^\|$ (Ar) (Eqs. 5 and 7) are more precise. Yet the calculation of $\Delta\phi_0^F$, $\Delta\phi_0^C$ from that requires the knowledge of $\sigma_{n,\infty}$. Unfortunately $\sigma_{n,\infty}$ is not available from the slope as the bulk is still *p*-type (see Section 1-1). We hence have to estimate $\sigma_{n,\infty}$ from $\sigma_{p,\infty}$ via

$$\sigma_{n,\infty} = F^2 u_n u_p K_B \qquad \text{(eq. 10)}$$

where $K_B$ is the mass action constant for band-band transfer. Literature gives $N_c = 7 \times 10^{18}$ cm$^{-3}$, $N_v = 2 \times 10^{18}$ cm$^{-3}$ [9,10]. We took the mobility values of 10 cm$^2$/(Vs) from literature[11,12]. For $\sigma_{n,\infty}$, we obtain $8.0 \times 10^{-18}$ Scm$^{-1}$. The evaluation leads to a value of about 740±60 mV for $\Delta\phi_0^C$ and 620±80 mV for $\Delta\phi_0^F$. This gives us an inversion iodine partial pressure of $P_{\text{minimum}} \approx 10^{-12}$ bar for the p to n inversion. Note that $P_{\text{minimum}} \propto c_{v,\infty}^{-2}$ and can hence shift to higher values if the material has lower impurity levels. From $\Delta\phi_0$ and $\lambda^*$, we derive Debye length a $\lambda = 2\pm1$ nm, from which we obtain $c_{v,\infty} \approx 10^{-6}$ mol/cm$^3$ (~1000 ppm, ~$10^{18}$ cm$^{-3}$). An acceptor impurity content of this magnitude is reasonable, considering the ubiquity of potential acceptor dopant (e.g. Na$_{Pb}^{/}$ or O$_{I}^{/}$ or O$_{i}^{//}$). Preliminary ICP results show Na concentration alone to be around 200 ppm (see SI section 4.2 for more details).

From $\sigma_{v,\infty} = 4 \times 10^{-8}$ Scm$^{-1}$ (Ar, dark, 40°C), we derive then a vacancy mobility of ~$10^{-8}$ cm$^2$/V·s.

## 1.2. Selective electrodes with severe space charge effects

Since we have indications that the contact Au-MAPI is influenced and under certain conditions even dominated by interfacial resistances[13] due to depletion of positive carriers (or possibly also due to charge transfer resistance), we investigate here the impact on the evaluation by assuming the extreme case of a dominant interfacial effect.

We show that these complications do not significantly influence the evaluation. In the galvanostatic experiment, we would then extract $R^{\perp}_{eon}$ from the steady sate and $R^{\perp}_{ion}$ from combining the steady state information with the external voltage response (analogous to the usual evaluation, but now yielding these boundary values, indicated by $\perp$, rather than the bulk values). The Mott-Schottky model gives then $R^{\perp}_j = F_j \cdot R_{j\infty}$ for the depleted carriers $h^{\bullet}$ and $V_I^{\bullet}$, where $F_j \propto (\lambda^*/d) exp(\pm \Delta\psi_0)/c_{\infty}$ ($\Delta\psi_0 = \Delta\phi_0 F/RT$) is composed of the normalized space charge potential in MAPI at the Au-contact and $d$ is the electrode distance. We can assume that the total adsorbed charge is not significant changed by $P(I_2)$ which only affects the minority level, i.e. $\lambda^* c_{v,\infty} \approx$ const. implying $\lambda^*$ ($c_{v\infty}$, $\Delta\psi_0$) and hence $\Delta\psi_0$ to be invariant. This shows that the partial pressure dependencies do not change. (A similar conclusion would be reached if the interfacial resistance is ascribed to charge transfer effects.)

The analysis given before is essentially based on the accumulation of $e^{\prime}$ under Ar. Under these conditions the interfacial effect could be assumed to be small, as Au tends to form an inversion layer, too. (Eqs. 1,5 of SI). The results for the thin films under depletion conditions should also be rather the same, as we used the first approximation, under which the depletion zone drops out of the conductance balance (Eqs 6,7 of SI). The same is expected for the interfacial resistance (MAPI-Au) as then in contact with $Al_2O_3$ or $TiO_2$ the depletion is augmented.

More problematic is the depletion evaluation for the composites (Eq. 9), as in that case the interfacial resistance can pretend too low a space charge potential. As we obtained a value of 300 mV as compared to 600 mV for the lower limit, the interfacial contribution would even lead to closer agreement.

## 1.3. Partial pressure dependence of the composite values

Here we also include the conductivity measurement for high values of $\varphi$. In spite of morphological problems yielding obvious conductivity blocking effects (*i.e.* $\beta_L \ll 1$) the iodine partial pressure dependence is even more reliable for high (rather than low $\varphi$ values) as long as the microstructure stays the same, since the space charge contribution to the measured value is much higher (see Fig. S2). From Refs.[2,14], we can give the calculated partial pressure dependence values in Table. S2.

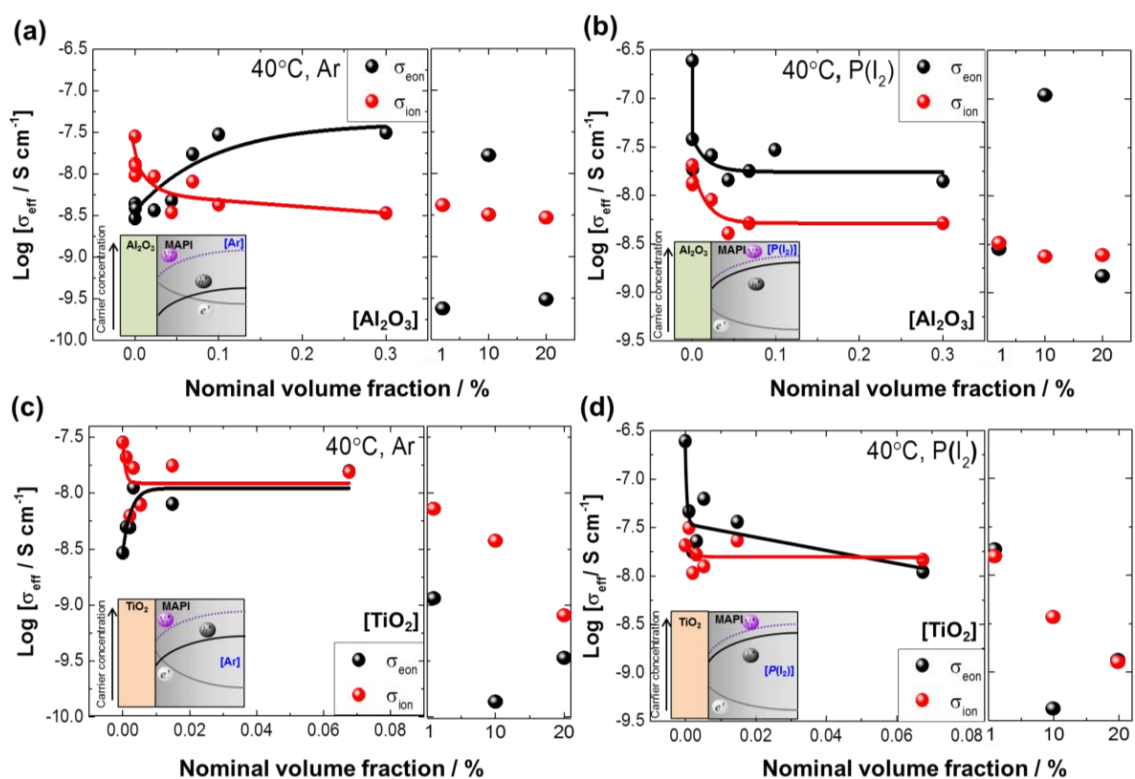

**Figure S2**. MAPI:Al$_2$O$_3$ (a) under Ar atmosphere (b) fixed iodine partial pressure (1x10$^{-6}$ bar) and MAPI:TiO$_2$ (c) under Ar atmosphere (d) under fixed iodine partial pressure. Electronic and ionic conductivities of composites as a function of volume fraction extracted from *d.c.* galvanostatic polarization at 40°C. Solid lines are guiding to the eye.

The measured values of iodine partial pressure variations in MAPI:Al$_2$O$_3$ and MAPI:TiO$_2$ composites are shown in Fig. S3 and Fig. S4, respectively. The observed slopes for MAPI:Al$_2$O$_3$ (see Fig. S3) for $\sigma_p$ and $\sigma_v$ (eon ~0.5, ion ~0) are all in accordance with the predictions from MS or GC models (see Table S2)[2,14]. This agreement is however only a consistency check as the slopes in the space charge models coincide with the bulk slopes as shown in Figs. S3 and S4. For the Ar-value, it is crucial to what $P(I_2)$ value we refer here. According to the theoretical slopes (1/4) and the experimental values for $\sigma_{p,\infty}$ we can approximately estimate $P(I_2)$ values $< 10^{-9}$ bar for Ar. The measurements in Fig. S3 and S4 may suggest a n-type bulk conductivity at this $P(I_2)$ value. This is indicated in Fig. S3(c) and Fig. S4(c) by the $\sigma_{eon}$ value in Ar being higher than the value at $P(I_2)=2\times10^{-7}$ bar.

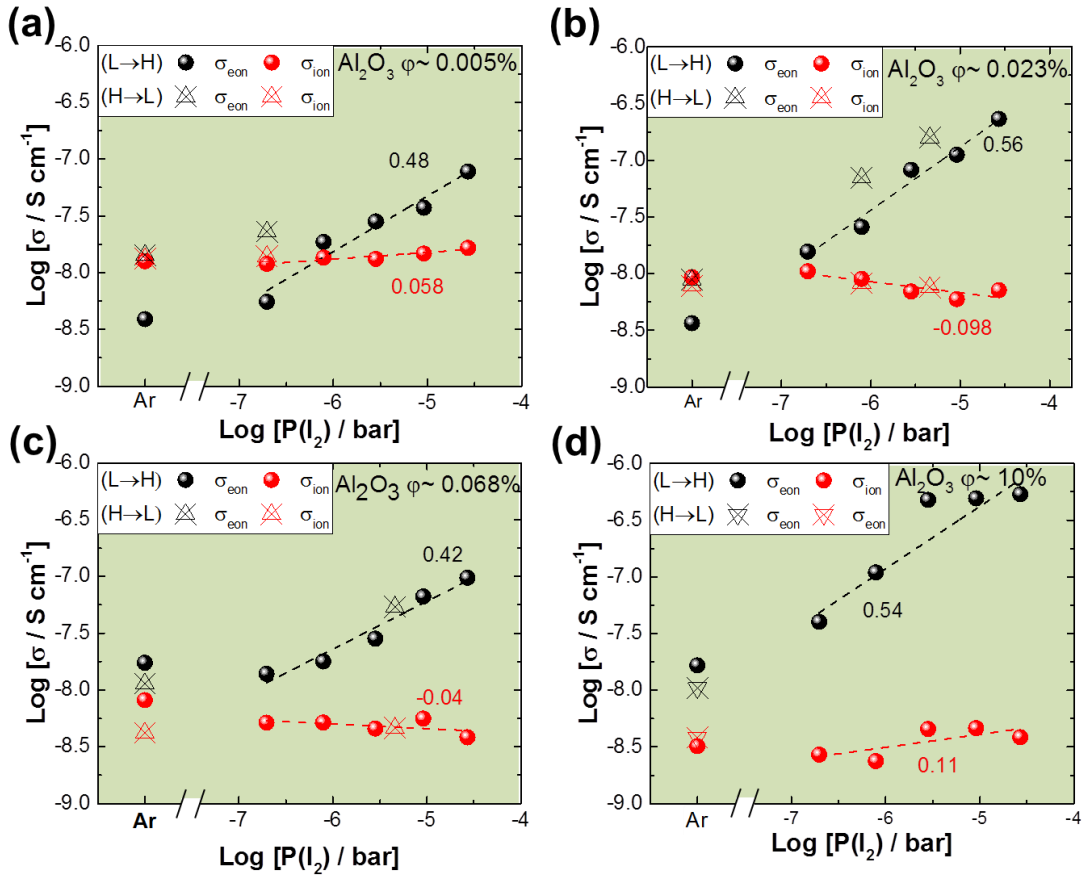

**Figure S3**. Iodine partial pressure dependence of the electronic and ionic conductivities of MAPI:$Al_2O_3$ composite film with different nominal volume fraction at 40°C. The numbers given are the slopes of the linear fit indicated by the dashed lines. The nominal volume fraction of samples is noted on the figure (L→H: from low to high iodine partial pressure, H→L: from high to low iodine partial pressure, carrier gas: Ar, black symbols: electronic conductivity, red symbols: ionic conductivity). As mentioned in the main text, the real oxide volume fractions in the samples is estimated (by ICP measurements, SI section 4) to be a factor of 10 higher than the nominal values.

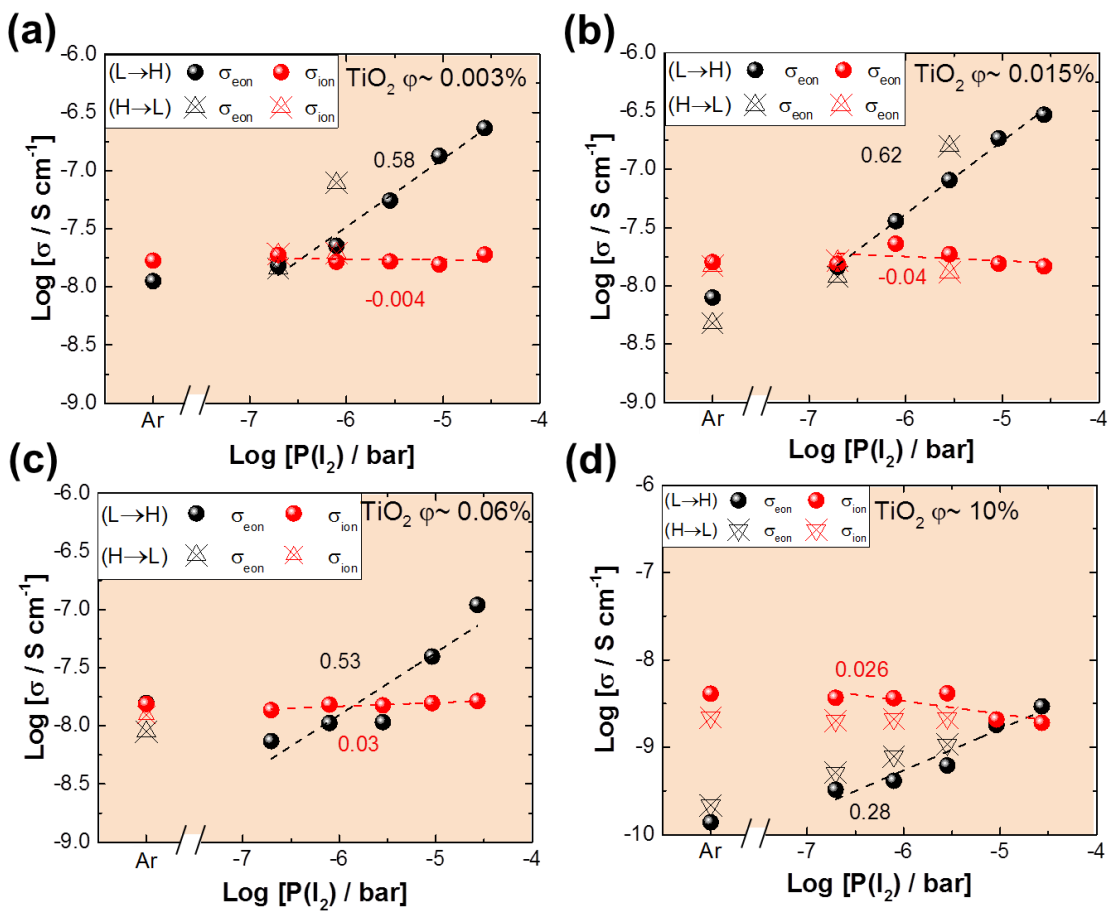

**Figure S4**. Iodine partial pressure dependence of the electronic and ionic conductivities of MAPI:TiO$_2$ composite film with different nominal volume fraction at 40°C. The numbers given are the slopes of the linear fit indicated by the dashed lines. The nominal volume fraction of samples is noted on the figure (L→H: from low to high iodine partial pressure, H→L: from high to low iodine partial pressure, carrier gas: Ar, black symbols: electronic conductivity, red symbols: ionic conductivity). As mentioned in the main text, the real oxide volume fractions in the samples is estimated (by ICP measurements, SI section 4) to be a factor of 10 higher than the nominal values.

**Table S2**. Iodine partial pressure dependencies of the excess conductivities in Mott-Schottky and Gouy-Chapman situations.

| | $N_\Delta^{\text{Mott-Schottky}}$ | $N_\Delta^{\text{Gouy-Chapman}}$ |
|---|---|---|
| v $\dfrac{\partial \ln(-\Delta\sigma_m)}{\partial \ln P_{I_2}}$ | $\sim 0$ | $-\dfrac{1}{2}N_{v\infty} \simeq 0$ |
| p $\dfrac{\partial \ln(-\Delta\sigma_m)}{\partial \ln P_{I_2}}$ | $N_{p\infty} - N_{v\infty} \simeq N_{p\infty} \simeq \dfrac{1}{2}$ | $N_{p\infty} - \dfrac{1}{2}N_{v\infty} \simeq N_{p\infty} \simeq \dfrac{1}{2}$ |
| n $\dfrac{\partial \ln(+\Delta\sigma_m)}{\partial \ln P_{I_2}}$ | $<0$ <br> $-\dfrac{1}{2}(v_0 \gg n_0)\ldots 0(n_0 \gg v_0)$ | $\dfrac{1}{2}N_{n,0} + \dfrac{1}{2}N_{n,\infty} - \dfrac{1}{2}N_{v,0} \simeq -\dfrac{1}{2}$ |

First layer of the space charge zone is denoted by subscript 0. $N_\Delta$ is the power in the power law $|\Delta\sigma_m| \propto P_{I_2}^{N_\Delta}$.

# 2. Conductivity under light

We performed *d.c.* galvanostatic conductivity experiments on both MAPI:$Al_2O_3$ and MAPI:$TiO_2$ composites with ion blocking electrodes on Au under 1 mW/cm$^2$ light illumination, and separate ionic conductivity and electronic conductivities. With increasing volume fraction, both electronic and ionic conductivity are reduced when $TiO_2$ (or $Al_2O_3$) particles are used as second phase. In spite of various unknowns, we observe depletion effects, even under conditions when we observed accumulation effects in the dark. The observed behavior suggest that the oxide particle significantly decrease the charge carrier lifetime of the photo-generated electronic carriers. The situation under light is complicated by the fact that also the ionic carrier concentration is enhanced[12], and probably a high trap density is involved. The increased is expected to lead to a pronouncedly weaker space charge potential than in the dark.

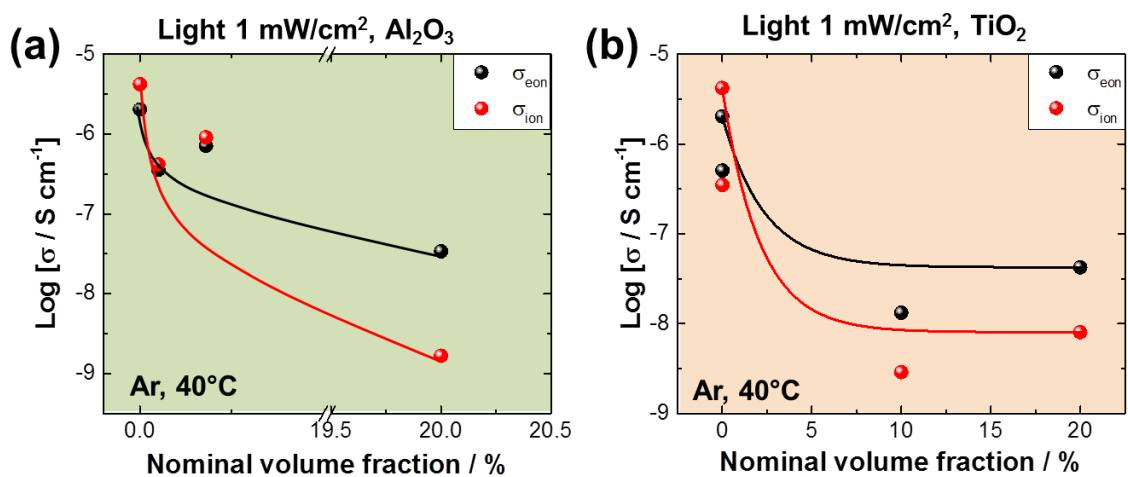

**Figure S5**. $\sigma_{eon}$ and $\sigma_{ion}$ of (a) MAPI:$Al_2O_3$ and (b) MAPI:$TiO_2$ composites as a function of volume fraction under Ar atmosphere. The data extracted from *d.c.* galvanostatic polarization at 40°C and 1 mW/cm$^2$ of light illumination (Xenon-arc lamp).

# 3. Indications of Pb adsorption

In this section we discuss the nature of the ion adsorption on $Al_2O_3$ and $TiO_2$ particles. We already experimentally confirmed a positive excess charge at the interface between MAPI and these oxides by conductivity measurements. In order to investigate the origin of this charge, we performed various experiments such as STEM-EDX, Zeta potential, ICP, NMR and XPS measurements. All the measurements are consistent with adsorption of $Pb^{2+}$ ions rather than $MA^+$ (or $I^-$) ions.

## 3.1. STEM and BET results

We performed SEM and STEM measurements to investigate the morphology of the composites characterized in this work. On the scale shown, we do not see particle aggregation (see Fig. S6). The particle size for both $Al_2O_3$ and $TiO_2$ is around 5-10 nm. The surface area of the nanoparticles is measured by Brunauer–Emmett–Teller (BET) $N_2$ adsorption isotherm. Consistently with the particle sizes, average surface areas of 115 $m^2/g$ ($Al_2O_3$) and 122 $m^2/g$ ($TiO_2$) are found. This surface area is large enough to explain the changes in conductivity in the oxide/hybrid perovskite composite (main text Fig. 2) even for the low nominal volume fractions (details in Fig. S11).

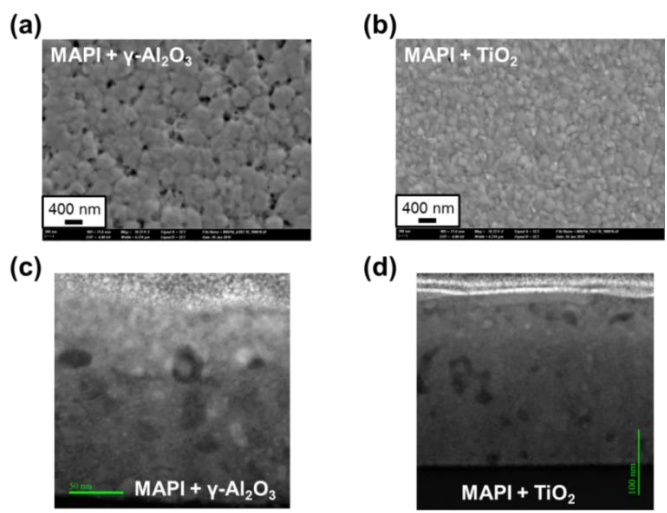

**Figure S6**. SEM images of the surface (a) MAPI:$Al_2O_3$ and (b) MAPI:$TiO_2$ composite thin films. STEM images of (c) MAPI:$Al_2O_3$ and (d) MAPI:$TiO_2$ composite thin-films.

## 3.2. Zeta potential measurement

We determined the zeta potential of $Al_2O_3$ and $TiO_2$ particles in various relevant solutions. The zeta potential of $TiO_2$ particles dispersed in DMSO containing 0.5 M $Pb(NO_3)_2$ is positive for all weight fractions, indicating a predominant adsorption of the $Pb^{2+}$ cation. The zeta potential of both $Al_2O_3$ and $TiO_2$ particles in DMSO contained 0.25 M KI + 0.25 M $Pb(NO_3)_2$ is also positive for all weight fractions, and indicates absence of a predominant I-adsorption. Quantitative conclusions are difficult to draw from such measurements; this also holds for the fact that the measured zeta potential values decrease with increasing particle weight fractions. The zeta potentials of both $Al_2O_3$ and $TiO_2$ particles dispersed in solutions containing MAI and KI salts show negative values suggesting adsorption of the $I^-$ anion if $Pb^{2+}$ is not present. The result also indicates that unlike $Pb^{2+}$, $MA^+$ adsorption is not significant (see also NMR results, Fig. S9).

The slightly negative values if the oxides are brought into contact with solutions of $PbI_2$ and $PbI_2$+MAI are not inconsistent with this pictures in view to the strong complexation of $Pb^{2+}$ by $I^-$ ($PbI_3^-$, $PbI_4^{2-}$ etc.)[15] in polar aprotic solvent. Consistently, EDX measurements showed clear Pb signals as well as weak I peaks (see Fig. S8).

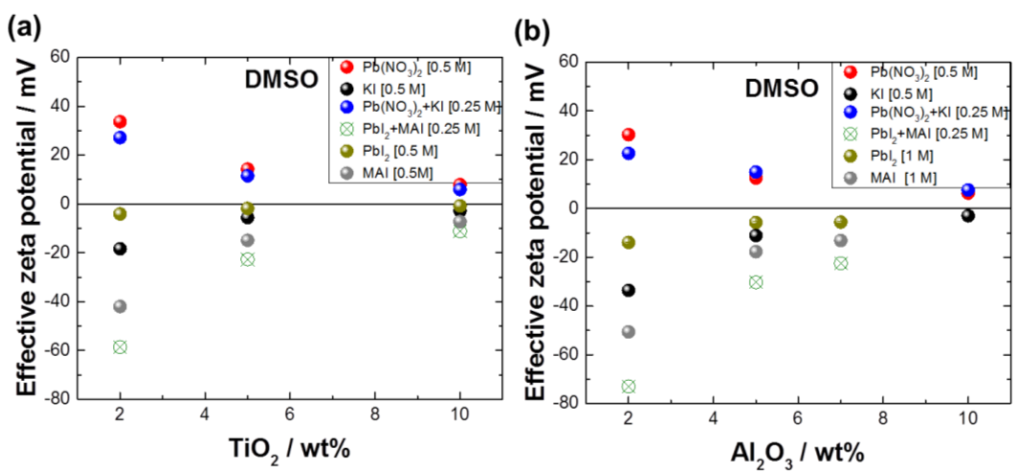

**Figure S7.** Effective zeta potential of (a) $Pb(NO_3)_2$, KI, $Pb(NO_3)_2$+KI, $PbI_2$+MAI, $PbI_2$ in DMSO as a function of $TiO_2$ particle weight fractions. Zeta potential of (b) $Pb(NO_3)_2$, KI, $Pb(NO_3)_2$+KI, $PbI_2$, $PbI_2$+MAI, MAI in DMSO as a function of $Al_2O_3$ particle weight fractions.

## 3.3. ICP and STEM-EDX results

In order to get insight into the amount of $Pb^{2+}$ adsorbed on the oxides surfaces, we immersed our oxide particles (~180 mg) into a solution of MAI and $PbI_2$ in DMSO. If Pb is adsorbed on the oxides surfaces, its concentration in DMSO should be decreased. We can quantify the Pb amount by ICP in a control solution (without nanoparticles) and compare it with the concentration left after adding $Al_2O_3$ or $TiO_2$ nanoparticles (and separating them by centrifuge). The control solution has also been centrifuged for better comparison. As expected, the amount of Pb found in solution is significantly lower when adding nanoparticles (see **Table. S3**). From the measured values, we can evaluate the Pb ion density per unit area on the particles and obtain Pb ion adsorption of approximately 30% of a monolayer. In this calculation a monolayer is defined by the extreme case of a close packing of $Pb^{2+}$ ions (i.e., based on the Pb ionic radius) on the oxide. For $\gamma$-$Al_2O_3$, a similar percentage is found when the definition is based on the square of the Al-Al distance. For $TiO_2$ this latter definition, (considering the square of the Ti-Ti distance in $TiO_2$) yields higher coverages (~70%.).

The situation of ion adsorption will be quantitatively very different between solution and solid case, but the qualitative conclusions are expected to be transferrable. The analysis of the particles by STEM-EDX is consistent with Pb and also I-accumulation on the surfaces (see Fig. S8). Consistent results are obtained by XPS (see Section 3.4). Combining this with the zeta-potential measurements (see Section 3.2), we can conclude that $Pb^{2+}$ is adsorbed and $I^-$ plays the role of the counter ion.

**Table S3**. Pb amount with and without immersion of insulating particles into a solution of MAI and $PbI_2$ in DMSO by ICP measurements. We dissolve MAI and $PbI_2$ salts in 2 mL of DMSO to obtain a 0.75 M solution, and add ~180 mg of $TiO_2$ or $Al_2O_3$ particles.

| Pb | Amount / $mgL^{-1}$ |
|---|---|
| Control solution (MAI+$PbI_2$ in DMSO, 0.75 M) | 301±5 |
| with ~180 mg $TiO_2$ nanoparticles | 266±2 |
| with ~180 mg $Al_2O_3$ nanoparticles | 265±5 |

*Note that the solutions were diluted 500 times before injection in the ICP.*

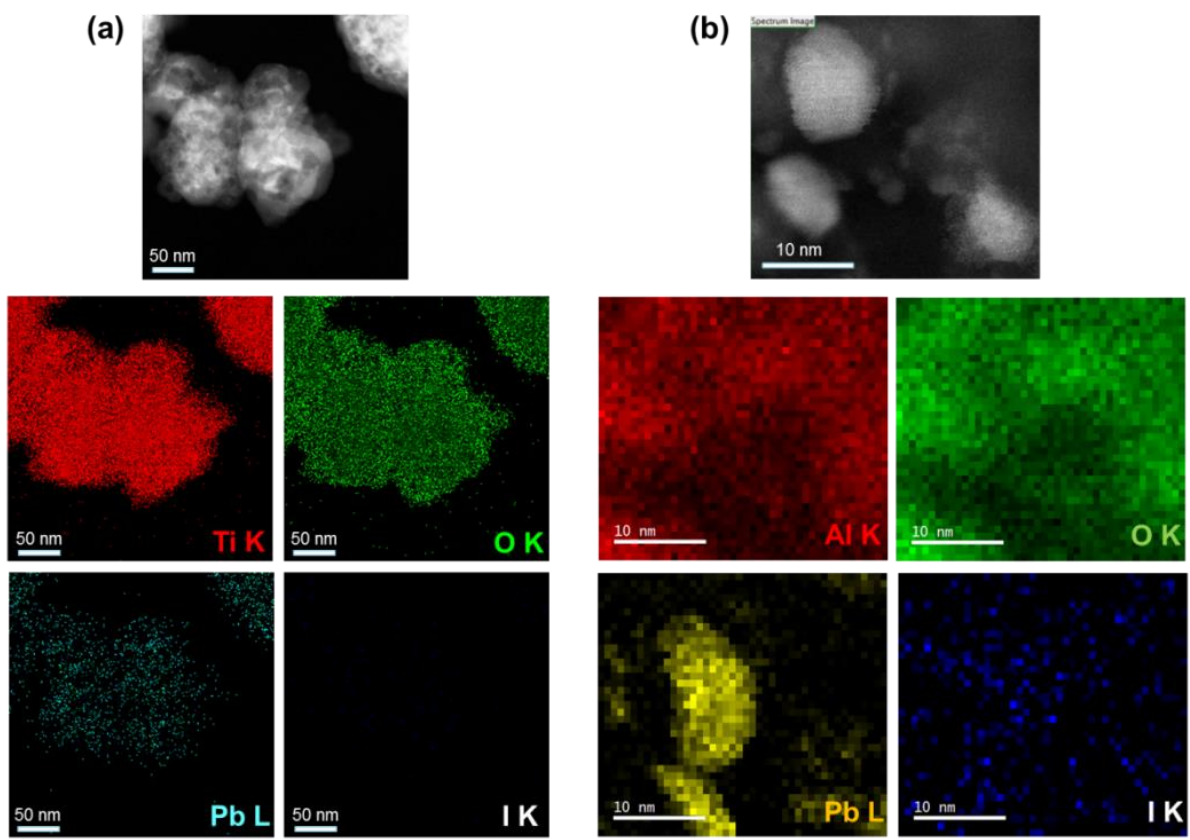

**Figure S8**. (a) STEM image and EDX color maps such as Ti, O, Pb, and I of immersed TiO$_2$ particles and (b) Al, O, Pb, and I of immersed Al$_2$O$_3$ particles.

## 3.4. NMR results

To obtain further evidence of ion adsorption at $Al_2O_3$ or $TiO_2$, we performed magic angle spinning (MAS) NMR spectroscopy experiments on the nanoparticles both pristine and after immersion in a MAI + $PbI_2$ in DMSO. The results do not show any sign of interaction between the oxides and $MA^+$ ions. In particular, we measured $^1H$ solid state NMR on MAPI:$Al_2O_3$ and MAPI:$TiO_2$, (and $^{27}Al$ NMR in $Al_2O_3$:MAPI composites) comparing different volume fractions to investigate the interaction of $MA^+$ with the insulating particles. As shown in Figs. S9, insulating admixtures do not affect the chemical shifts at any given volume fraction in $^1H$ and $^{27}Al$ NMR spectra.

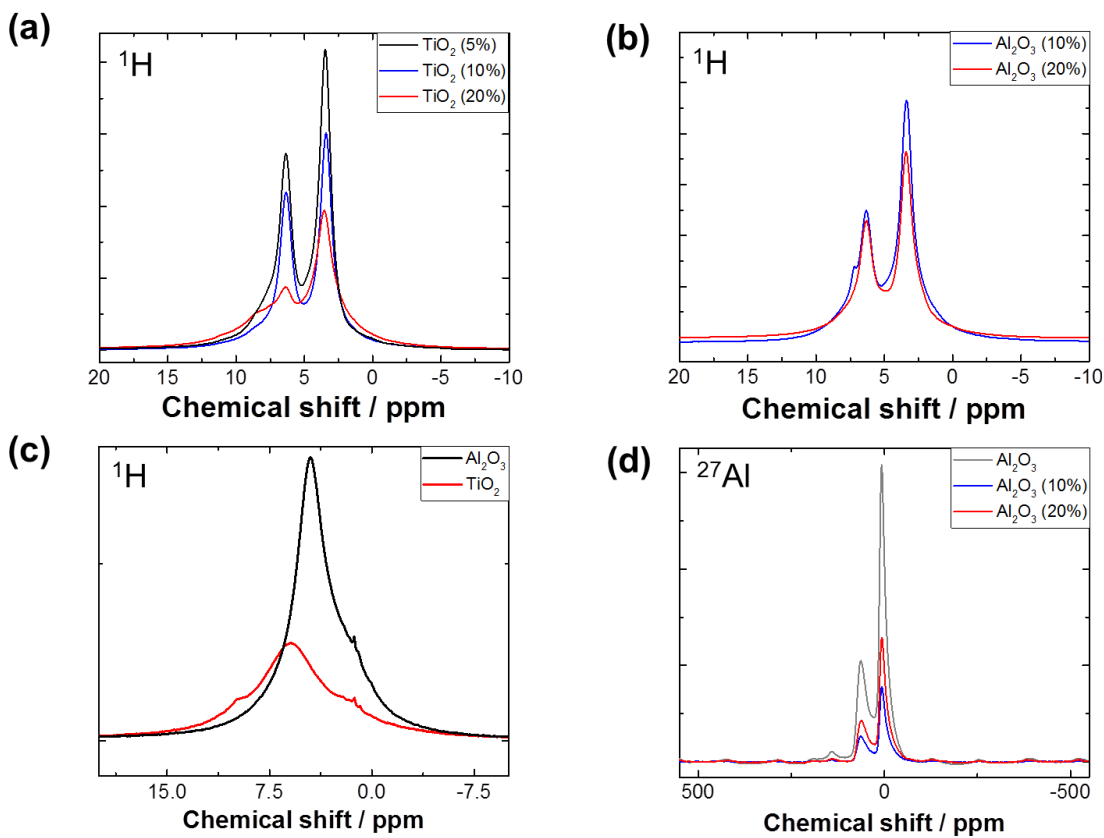

**Figure S9.** $^1H$ MAS NMR spectra of (a) $TiO_2$ (b) $Al_2O_3$ powders after immersion in MAI and $PbI_2$ contained DMSO as a function with different volume fractions (c) $^1H$ MAS NMR spectra of pristine $Al_2O_3$ and $TiO_2$ powders. (d) $^{27}Al$ MAS NMR spectra in MAPI:$Al_2O_3$ composite and pure $Al_2O_3$ powder. The peak corresponds to water and hydroxide groups adsorbed on the oxide surface. All the measurements are done at room temperature.

## 3.5. XPS results

We performed XPS measurements on $TiO_2$ and $Al_2O_3$ powders immersed in a solution of MAI and $PbI_2$ in DMSO (and subsequently dried under vacuum), and also on MAPI:$TiO_2$ and MAPI:$Al_2O_3$ composite films. The powder samples were pressed on indium foil for the measurements. Samples were transferred from the glovebox to the XPS instrument under Ar atmosphere to avoid contamination from air. Considering the powders immersed in solution, we observed weak Pb and I signals for both $Al_2O_3$ and $TiO_2$ as shown in Figs. S10 (c), (d), (g), and (h). In the case of $Al_2O_3$, Pb and I signals are found stronger than for $TiO_2$. Both pristine powders (before immersion) do not show any Pb or I signals. This observation clearly indicates adsorption of Pb and I on the oxide particles. Focusing on the carbon signal, C-O-Ti, C-O[16-18] and C-C (or C-H) peaks appeared in pristine $TiO_2$ powders, and are present also after immersion in the MAI/$PbI_2$ solution. Pristine $Al_2O_3$ powders also show a C-C (or C-H) and C-O peaks, which again are preserved after immersion. As expected, both pure MAPI films and composite MAPI:$TiO_2$ and MAPI:$Al_2O_3$ films show pronounced peaks referring to the C-N, C-C and C-H bonds (~286 eV and 284 eV). Notably, the C-N peak is not visible in the $TiO_2$ or $Al_2O_3$ powders after immersion in the MAI/$PbI_2$ solution. This observation is strongly corroborated by the total absence of a nitrogen signal (Fig. S10 (b),(f)) in those samples, in striking contrast with the MAPI:$TiO_2$ and MAPI:$Al_2O$ composite films. This clearly indicates negligible adsorption of MA cations on the oxide powders. Combining this evidence with the zeta potential measurements, we can conclude that the surfaces of $TiO_2$ and $Al_2O_3$ adsorb preferentially $Pb^{2+}$ cations, with $I^-$ as counter ions.

We also note the presence of a shift of Pb and I peaks to higher binding energies possibly due to interaction with the indium foil. More evidences are required to clarify this aspect.

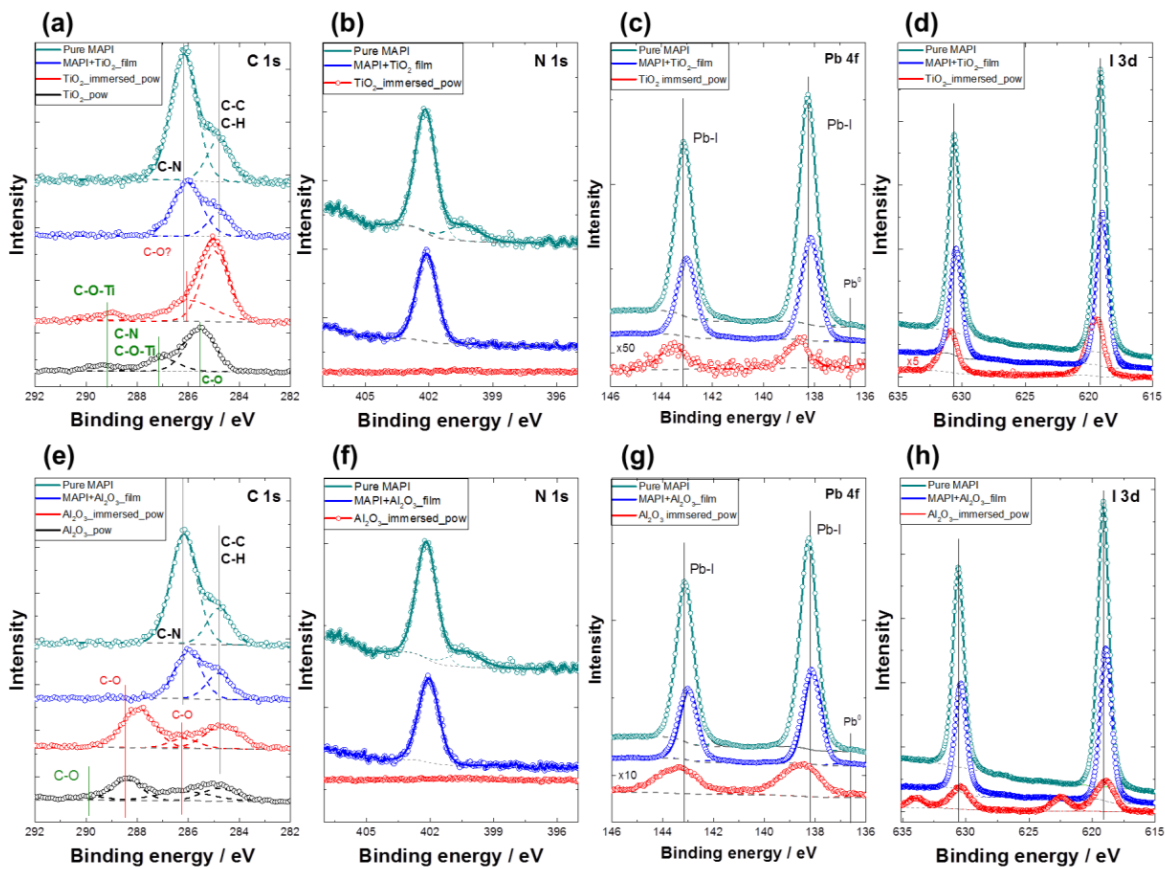

**Figure S10.** XPS measurements of TiO$_2$ powders (a)-(d) and Al$_2$O$_3$ powders (e)-(h) with (red line) and without (black line) immersion. Blue lines indicate MAPI:TiO$_2$ or MAPI:Al$_2$O$_3$ composite films (analogous to the ones studied in the main text). For comparison, pristine MAPI signal is shown by the cyan line. Circles correspond to raw data, while the thin dashed lines indicate the Shirley background and the fits to the individual peaks and the sum of the fits is displayed by the thin continuous line. Binding energy windows refer to (a), (e) C 1s, (b), (f) N 1s, (c), (g) Pb 4f, and (d), (h) I 3d.

# 4. Chemical analyses

We already discussed the quantification of Pb adsorption on oxide particles by ICP (SI Section 3.3.). Here we apply chemical analyses to assess the true oxide content in MAPI:oxide composites, and also to determine the impurity content in MAPI films.

## 4.1. True oxide volume fraction in MAPI:TiO$_2$ composites

We use chemical analyses (ICP and STEM-EDX) to check the true oxide content in the composite MAPI:TiO$_2$ films. The incentive for these measurements lies in the fact that we find significant conductivity effects (see Fig 2 main text) already at very low volume fractions, possibly indicating a particle decoration that comprises almost the entire film. This occurrence appears to be favored by accumulation of the oxide nanoparticles at the Au electrodes during the spin-coating process. Indeed, STEM-EDX measurements (Fig. S11) shows that even for a low nominal volume fraction (1%) a substantial amount of TiO$_2$ nanoparticle is found, which seems to form a percolating path between the Au electrodes. Consistently, ICP measurements performed after entirely dissolving a MAPI:TiO$_2$ composite with a nominal TiO$_2$ content of 1% (analogous to the ones used in the main text for electrical experiments) show a significant Ti content, corresponding to a real volume fraction which is 10 times higher (~10%). This can be ascribed to interactions between oxide and Au (or sapphire substrate) and possibly also to precipitation of the nanoparticles during the spin-coating step. Of course, at very high $\varphi$-values true and nominal values converge.

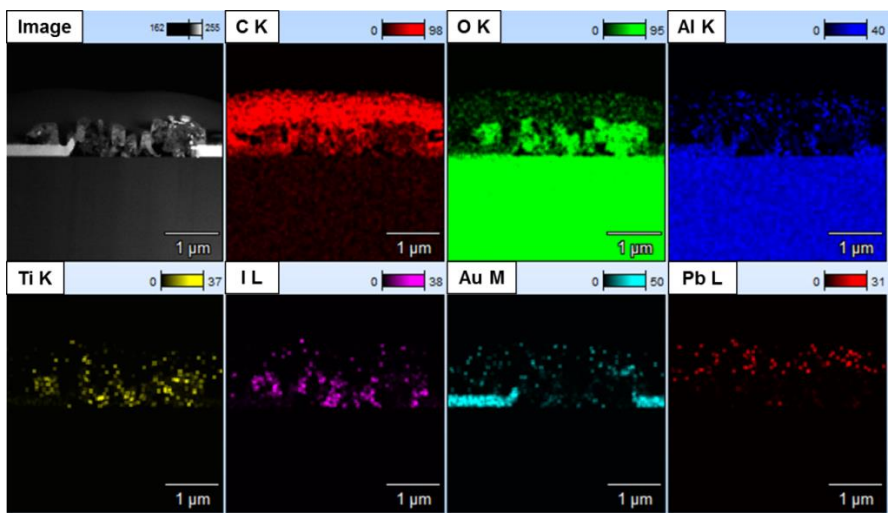

**Figure S11.** STEM-EDX images of MAPI and TiO$_2$ composite (nominal φ in solution ~1%, particle size 5-10 nm) with Au electrodes on Al$_2$O$_3$ substrates. Carbon is used as coating layer on the film. In the middle of the Au electrodes, we observed an arrangement of apparently percolating TiO$_2$ particles. Our measurements suggest that this effect occurs even at low volume contribution.

## 4.2. Impurity level

It is of interest to assess the impurity level in MAPI, particularly considering that both Na and O, which are ubiquitous, can act as acceptor dopants. Unfortunately, quantifying the real bulk oxygen content is extremely challenging, due to the presence of large background oxygen quantities (e.g. from gas incorporation/adsorption or from solvent such as DMSO) which may however not act as dopants. We focus therefore on Na, which can be quantified by ICP, showing an impurity content of 200 ppm in the precursor solution used to spin-coat the MAPI films. This value is consistent with the total bulk impurity concentration calculated from electrical measurements (~1000 ppm, see SI 1.1.4), indicating that the defect situation in MAPI at room temperature is extrinsic rather than intrinsic. Further experiments are needed to confirm this claim, also in view of the fact that the total Na content (as measured by ICP in the precursor) does not necessarily correspond with the electrically active impurity content in solid MAPI thin films.